\documentclass{aa}  
\usepackage{threeparttable}
\usepackage{natbib}
\usepackage{xcolor}
\usepackage{placeins}

\usepackage{microtype}
\usepackage{hyperref}
\usepackage{graphicx}
\usepackage{txfonts}
\PassOptionsToPackage{hyphens}{url}
\usepackage{hyperref}
\begin{document}

\title{Metal-poor nuclear star clusters in two dwarf galaxies near Centaurus A suggesting formation from the in-spiraling of globular clusters{\thanks{Based on observations collected at the European Southern Observatory under ESO programmes 0101.A-0193(B), 084.D-0818(C), and 069.D-0169(A).}}}

\titlerunning{Metal-poor nuclear star clusters in dwarf galaxies near Cen A}

   \author{Katja Fahrion\inst{1}
          \and
          Oliver M\"uller\inst{2}
          \and 
          Marina Rejkuba\inst{1}
          \and
          Michael Hilker\inst{1}
          \and
          Mariya Lyubenova\inst{1}
          \and
          Glenn van de Ven\inst{3}
          \and\\
          Iskren Y. Georgiev\inst{4}
          \and
          Federico Lelli\inst{1}
          \and
          Marcel S. Pawlowski\inst{5}
          \and
          Helmut Jerjen\inst{6}
          }

   \institute{European Southern Observatory, Karl Schwarzschild Stra\ss{}e 2, 85748 Garching bei M\"unchen, Germany\\
   				\email{kfahrion@eso.org}
         \and %2
             Observatoire Astronomique de Strasbourg  (ObAS),
Universite de Strasbourg - CNRS, UMR 7550 Strasbourg, France
        \and 
        Department of Astrophysics, University of Vienna, T\"urkenschanzstrasse 17, 1180 Wien, Austria
        \and
        Max-Planck-Institut f\"{u}r Astronomie, K\"{o}nigstuhl 17, 69117 Heidelberg, Germany
        \and
        Leibniz-Institut f\"ur Astrophysik Potsdam (AIP), An der Sternwarte 16, D-14482 Potsdam, Germany
             \and
 Research School of Astronomy and Astrophysics, Australian National University, Canberra, ACT 2611, Australia
             }
   \date{}

% 5 {} token are mandatory
 
  \abstract{Studies of nucleated dwarf galaxies can constrain the scenarios for the formation and evolution of nuclear star clusters (NSC) in low-mass galaxies and give us insights on the origin of ultra compact dwarf galaxies (UCDs). We report the discovery of a NSC in the dwarf galaxy KKs58 and investigate its properties together with those of another NSC in KK197. Both NSCs are hosted by dwarf elliptical galaxies of the Centaurus group. Combining ESO VLT MUSE data with photometry from VLT FORS2, CTIO Blanco DECam, and HST ACS, as well as high-resolution spectroscopy from VLT UVES, we analyse the photometric, kinematic and stellar population properties of the NSCs and their host galaxies. We confirm membership of the NSCs based on their radial velocities and location close to the galaxy centres. We also confirm the membership of two globular clusters (GCs) and detect oblate rotation in the main body of KK197. Based on high signal-to-noise spectra taken with MUSE of the NSCs of both KKs58 and KK197 we measure low metallicities, \mbox{[Fe/H] = $-1.75 \pm 0.06$ dex} and \mbox{[Fe/H] = $-1.84 \pm 0.05$ dex}, and stellar masses of $7.3 \times 10^5 M_\sun$ and $1.0 \times 10^6 M_\sun$, respectively. Both NSCs are more metal-poor than their hosts that have metallicities of $-1.35 \pm 0.23$ dex (KKs58) and $-0.84 \pm 0.12$ dex (KK197). This can be interpreted as NSC formation via the in-spiral of GCs.
 The masses, sizes and metallicities of the two NSCs place them among other NSCs, but also among the known UCDs of the Centaurus group. This indicates that NSCs might constitute the progenitors of a part of the low-mass UCDs, although their properties are almost indistinguishable from typical GCs.}
   \keywords{galaxies: kinematics and dynamics -- galaxies: dwarf -- 
            galaxies: star clusters: general}
   \maketitle
%
%-------------------------------------------------------------------

\section{Introduction}
\label{sect:intro}
Dwarf galaxies are the most abundant type of galaxies in the Universe. Typically, they are classified as either dwarf irregulars or dwarf ellipticals based on their optical morphology, the presence of ongoing or recent star formation, or gas content \citep{Mateo1998} although this is a simplified and incomplete scheme \citep[see][for a more detailed discussion about different dwarf galaxy types]{Tolstoy2009, Ivkovich2019}. While dwarf irregulars have
a relatively high gas content and exhibit ongoing star formation (e.g \citealt{Warren2004, Warren2006, McGaugh2017}), the gas content of dwarf ellipticals is usually very low (e.g. \citealt{Spekkens2014}). Dwarf ellipticals can be further subdivided into nucleated and non-nucleated depending on the presence of a nuclear star cluster (NSC) in or near the photometric centre of the dwarf galaxy (e.g. \citealt{Ordenes2018}). However, bright star clusters at central positions have also been found in dwarf irregulars (e.g. \citealt{Georgiev2009a}).

With effective radii of $5 - 10$ pc and stellar masses \mbox{$M_\ast \approx 10^4 - 10^7 M_\sun$}, NSCs are typically more massive and denser than globular clusters (GCs, e.g. \citealt{Boker2004, Walcher2005, Cote2006, Turner2012, Puzia2014, Georgiev2016, Spengler2017}). They can have complex dynamics \citep{Seth2010, Lyubenova2013, Lyubenova2019, Fahrion2019b}, broad metallicity distributions \citep{Spengler2017}, and multiple stellar populations \citep{Walcher2006, Lyubenova2013, Kacharov2018}. They often demonstrate a significant contribution from young populations, particularly in NSCs of late-type galaxies \citep{Rossa2006, PuziaSharina2008, Paudel2011, Feldmeier-Krause2015}.

Although NSCs are very common in all types of galaxies (e.g. \citealt{Georgiev2009b, Eigenthaler2018}), their formation is still under debate and two main pathways are usually discussed. In the GC accretion or GC in-spiral scenario, the NSC builds out of the (dry) mergers of GCs that migrate into the galaxy's centre due to dynamical friction (e.g. \citealt{Tremaine1975, CapuzzoDolcetta1993, CapuzzoDolcetta2008, Agarwal2011, ArcaSedda2014, Hartmann2011}). In this scenario, the NSC is expected to reflect the metallicity distribution of the coalesced GCs \citep{Perets2014}, but if gas-rich or young star clusters merged, the resulting NSC can show a complex star formation history \citep{Antonini2014, Guillard2016}. 
Contrary, in the in-situ formation scenario, the NSC forms independently from the GC system directly at the centre of the host galaxy from accumulated, dense gas \citep{MihosHernquist1994, Bekki2006, Bekki2007, Antonini2015}. The efficiency of the latter scenario strongly depends on mechanisms that funnel gas to the centre (e.g. \citealt{Milosavljevic2004, Schinnerer2008}).

In reality, most likely a mixture of both processes occurs \citep{DaRocha2011, Spengler2017, Ordenes2018, Sills2019}, but the relative contributions of both formation channels might depend on the host's mass. 
Studies of early-type galaxies in the Fornax and Virgo galaxy clusters have suggested that the in-situ formation might be the dominant channel of NSC formation in massive galaxies ($> 10^9 M_\sun$), while the GC accretion is more efficient in low mass galaxies because of shorter dynamical friction timescales (e.g. \citealt{Lotz2004, Cote2006, Turner2012, SanchezJanssen2019b}).
The proposed NSC formation scenarios and their mass dependence also seems to be reflected in the nucleation fraction of galaxies. Within the Next Generation Virgo Cluster Survey (NGVS, \citealt{Ferrarese2012}), \cite{SanchezJanssen2019} found a peak of the nucleation fraction of $> 90\%$ at $M_\ast \sim 10^9 M_\sun$. In more massive galaxies, co-existence of a NSC with a central super massive black hole (SMBH) can result in the dissolution of the NSC if the SMBH is massive enough (e.g. \citealt{Antonini2013}). At the low mass end, the fraction of nucleated dwarf galaxies drops, possibly because NSCs are vulnerable to environmental effects and feedback in the shallow central gravitational potential.

In the context of galaxy assembly and evolution, it is often suggested that the remnant nuclei of disrupted galaxies may be the progenitors of the most massive GCs (e.g. \citealt{Zinnecker1988}) and of ultra compact dwarf galaxies (UCDs, e.g. \citealt{Phillipps2001, Drinkwater2003, Pfeffer2013, Strader2013}). In the Milky Way (MW), the GC M\,54 is perhaps the most prominent case as it is still located at the centre of the Sagittarius dwarf galaxy \citep{Ibata1997, Bellazzini2008, Sills2019} and thus can be clearly classified as a NSC. But also $\omega$Centauri, the most massive MW GC, is often interpreted as the remnant NSC of a disrupted galaxy (e.g. \citealt{Hilker2000, King2012, Ibata2019, Milone2019}).

Nucleated dwarf galaxies have been studied in various environments, from dense galaxy clusters such as Virgo, Coma or Fornax to less populated groups (e.g. \citealt{Georgiev2009a, Georgiev2014, denBrok2014, Ordenes2018}).
In this work, we report on two nucleated dwarf elliptical galaxies that are confirmed members of the Centaurus group \citep{Mueller2019}. 
The two large galaxies in the Centaurus group are Cen\,A (NGC\,5128) at a distance of $\sim$ 3.8 Mpc \citep{Rejkuba2004}, and M\,83 (NGC\,5236) at $D \approx$ 4.8 Mpc \citep{Herrmann2008, Radburn-Smith2011}.
In the first survey of dwarf elliptical galaxies in the Centaurus group, \citet{Jerjen2000b} identified 13 potential members and established group membership for five dwarf ellipticals through surface brightness fluctuation and velocity measurements. \citet{Jerjen2000} pointed out potential nuclei in the two dwarf ellipticals ESO 219-010 and ESO 269-066, although a chance projection of a (nearby) star on the galaxy centre could not be excluded. Using observations with the Advanced Camera for Surveys (ACS) onboard the \textit{Hubble Space Telescope} (HST), \cite{Georgiev2009b} studied the properties of NSC and GC candidates in nearby dwarf galaxies, including 24 dwarfs in the Centaurus group complex, suggesting that four of them have NSC candidates (ESO 059-01, ESO 223-09, ESO 269-66 and KK197, \citealt{Georgiev2009a}). 
Recently, several surveys have targeted the Centaurus group providing a more complete census of the dwarf galaxy system (\citealt{Crnojevic2014, Crnojevic2016, Crnojevic2019, Muller2015, Muller2017, Mueller2018, Mueller2019, Taylor2016, Taylor2018}) that as of now includes $\sim$70 dwarf galaxy candidates.

In this paper, we present the analysis of the two nucleated dwarf galaxies KKs58 and KK197 in the Centaurus group, based on data taken with the Multi Unit Spectroscopic Explorer (MUSE). 
As mentioned, KK197 was among the dwarf galaxies previously analysed with HST data by \cite{Georgiev2009a}, and photometric properties of its NSC and two GC candidates were presented in \cite{Georgiev2009b}. 
We report the discovery of a new NSC in KKs58 (also known as Centaurus A-dE3, \citealt{Jerjen2000}), which is the brightest dwarf galaxy in the sample of nine dwarf galaxies that were recently confirmed Cen\,A group members using ESO Very Large Telescope (VLT) FORS2 data \citep{Mueller2019}. In a spectroscopic follow-up with MUSE, under observing program 0101.A-0193(B), we measure radial velocities of a selection of these galaxies to investigate the velocity distribution of the dwarf satellites arranged in a plane around Centaurus A \citep{Tully2015,Mueller2016,Mueller2018Sci}. Upon inspection of MUSE data we discovered that the bright compact source located in the centre of KKs58 has a velocity in agreement with the host galaxy, and thus it constitutes the newly discovered NSC of KKs58. Additionally, we could confirm the star cluster candidate KK197-02 from \cite{Georgiev2009a} to be the NSC of KK197.
In the following section, we describe the MUSE data. Section \ref{sect:photometry} describes the DECam and HST photometry, respectively, and our MUSE analysis is detailed in Sect. \ref{sect:spect}. In Sect.~\ref{sect:spect}, we also present a measurement of dynamical mass of the NSC belonging to KK197 based on the high-resolution UVES spectrum. We discuss our results in Sect. \ref{sect:discussion} and conclusions in Sect. \ref{sect:conclusions}.

\begin{table}
    \centering
    \caption{Basic information about KKs58 \citep{Mueller2019} and KK197 \citep{Sharina2008, Georgiev2009a}.}
    \begin{threeparttable}
    \begin{tabular}{ l  c c }\hline \hline
    Property & KKs58 & KK197  \\ \hline
  RA (J2000)	&	13:46:00.8 & 13:22:01.8\\
  DEC  (J2000) &  $-$36:19:44  & $-$42:32:08\\
 D (Mpc) & 3.36 $^{+0.18}_{-0.02}$  & 3.98  \\
 d$_\text{3D, Cen\,A}$ (kpc) & 574 & 55  \\
 M$_V$ (mag) & $-11.93^{+0.12}_{-0.01}$ & $-$13.04  \\
 %$r_\text{eff}$ (pc) & $430^{+23}_{-2}$ & 733   \\
 \hline
\end{tabular}
    \label{tab:KKs58}
    \end{threeparttable}
\end{table}
%--------------------------------------------------------------------

\begin{figure*}
\centering\includegraphics[width=0.49\textwidth]{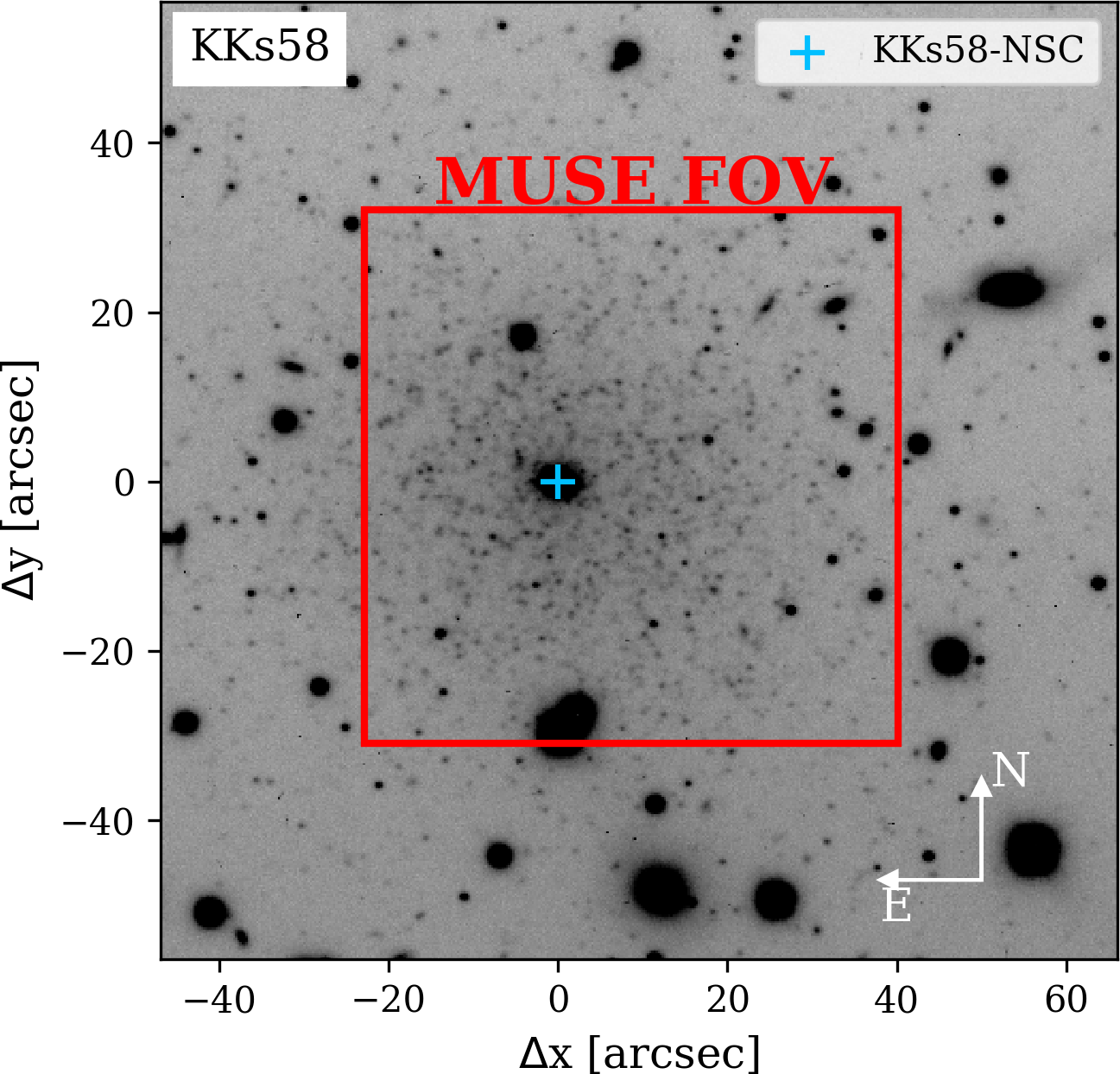}
\includegraphics[width=0.49\textwidth]{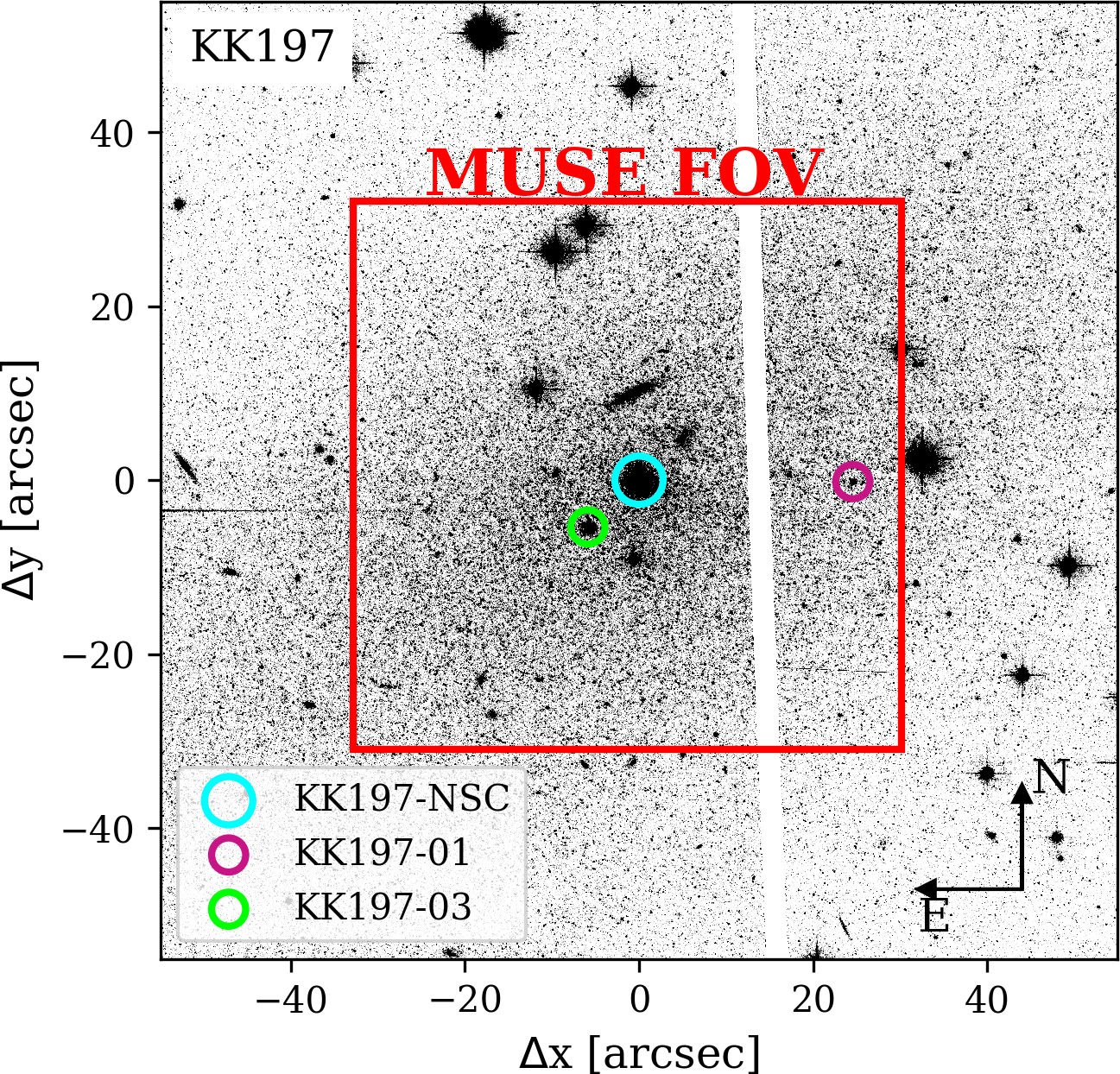}
\caption{Cutout of the FORS2 $I$-band image of KKs58 (left, 10\arcsec $\approx$ 160 pc) and of the HST F606W image of KK197 (right, 10\arcsec $\approx$ 190 pc). The pointings of the MUSE FOV (1$\times$1 arcmin $\approx$ 1$\times$1 kpc) are indicated by the red squares. North is up and East to the left, the $xy$ coordinates are relative to the NSCs. We indicate the positions of KKs58-NSC and the star cluster candidates of KK197 \citep{Georgiev2009a}.}
\label{fig:MUSE_and_FORS}
\end{figure*}

% --------------------------- MUSE DATA -------------------------

\section{MUSE data}
\label{sect:data}
The MUSE instrument is an integral field spectrograph mounted at UT4 on the Very Large Telescope on Paranal, Chile \citep{Bacon2010}. For our observations, we used the so-called Wide Field Mode (WFM) of MUSE, which has a field-of-view (FOV) of 1\arcmin$\times$1\arcmin\, sampled at $0\farcs2$~pix$^{-1}$, and requested observations in Service Mode under so-called filler conditions (i.e.\ relaxed seeing and thin sky transparency) because we were interested in measuring radial velocity from spatially integrated spectra.
Along the spectral dimension, in the adopted `nominal' setup, MUSE covers the optical wavelength range from 4500 to 9300\,\AA\,with a spectral resolution of $\sim$ 2.5\,\AA\,at 7000\,\AA\,sampled at 1.25\,\AA\,pix$^{-1}$.

Observations of KKs58 were acquired on 9 May 2018 under below-average sky conditions for Paranal (thin clouds and about $1\farcs2$ seeing). The total exposure time on target was 2000 seconds. The observing strategy included four science exposures, each 500 seconds long, interleaved with two offset sky exposures of 250 seconds, in a sequence {\it OSOOSO}, where \emph{'O'} is object and \emph{'S'} is sky exposure. The offset sky location was selected 72\arcsec\, East and 70\arcsec\, North of the central pointing. We implemented small dithering offsets of 1--2\arcsec, and 90 degree rotation between subsequent exposures in order to minimise the noise residuals due to slicers and different channels.  

The MUSE observations of KK197 were carried out within two observation blocks (OBs), each with a single {\it 'OSO'} sequence of $2 \times 1160$~sec science exposures on target bracketing a single 580 sec long exposure of an adjacent empty sky region. The first OB was taken on 16 April 2018 under mostly clear sky. The atmospheric conditions during the night of 8-9 May were worse, with initially some thick clouds affecting the observation, later improving to thin clouds. Therefore, to compensate for loss of sensitivity due to clouds, part of the observing sequence was repeated resulting in 3 on-target exposures \emph{'O'} of $1160$~sec each interleaved with $580$~sec long offset sky exposures \emph{'O'} in a \emph{'OSOOS'} sequence. Hence, the total exposure time for KK197 amounts to 5800~seconds. As for KKs58, small dithers and 90 degree rotation were implemented between science exposures. The fifth, repeated, science exposure had the same offset and rotator position (PA = 180$^\circ$) as the first science exposure taken on 16 April.

We downloaded the MUSE Internal Data Products from the ESO Science Archive that included data reduction based on the MUSE pipeline version 2.2 \citep{Weilbacher2012}. The data have been pre-processed, bias and flat-field corrected, astrometrically calibrated, sky-subtracted, wavelength calibrated and flux calibrated \citep{Hanuschik2017}\footnote{See also \url{http://www.eso.org/observing/dfo/quality/PHOENIX/MUSE/processing.html}}. The sky subtraction used the offset sky exposures.
To further reduce the sky residual lines, we applied the Zurich Atmosphere Purge (ZAP) principal component analysis algorithm \citep{Soto2016}. Figure~\ref{fig:MUSE_and_FORS} shows portions of the deep I-band stacked FORS2 image of KKs58 \citep{Mueller2019} and the HST ACS image of KK197 \citep{Georgiev2009a} with the MUSE FOVs indicated in red. These images were used to guide the selection of 'empty sky' regions with no evident stars or galaxies to be used by ZAP. The optimal sky mask turned out to be a selection of the faintest 9 - 12\% pixels on the white light image created from the MUSE cube. With the dedicated sky observations and ZAP, bright sky emission lines could be reduced sufficiently, but in low S/N spectra, telluric lines above $\sim$ 7500 \AA\,remain. As described in Sect. \ref{sect:ppxf}, we do not use the spectra above 7100 \AA.

Using DAOPHOT \citep{Stetson1987}, we measured the full width at half maximum (FWHM) of the point spread function (PSF) in the collapsed MUSE images, as described in \citet{Mueller2018}. Both MUSE cubes have an image quality of $1 \farcs 0$. 

%--------------------------------------------------------------------
%--------------------------------------------------------------------%--------------------------------------------------------------------

\section{Photometry}
\label{sect:photometry}
In the following, we describe the photometric properties of KK197, KKs58 and their star clusters. KK197 was previously studied by \cite{Georgiev2009a} using HST data and for KKs58, we use \cite{Muller2015} DECam data.

\subsection{KKs58 DECam photometry}
\label{sect:DECam_phot}
KKs58 was observed with FORS2 and photometry of the individual bright red giant branch (RGB) stars was used to measure the distance to the galaxy using the tip magnitude of the RGB \citep{Mueller2019}. The bright source located approximately in the centre of the galaxy (at coordinates ($\Delta y, \Delta x$) = (0, 0) in Fig.~\ref{fig:MUSE_and_FORS}) was saturated on individual FORS2 images. 
To study the photometric structure of KKs58 and its NSC in more detail, we used the $g$-and $r$-band DECam images from \citet{Muller2015}.
The data have an exposure time of 120 seconds in both filters. Using DAOPHOT aperture photometry \citep{Stetson1987}, we obtained extinction-corrected magnitudes of the NSC of $g_\text{NSC} = 18.44 \pm 0.05$ mag and $r_\text{NSC}= 17.90 \pm 0.06$ mag, respectively, using galactic extinction values in the $g$ and $r$-band of 0.200 and 0.134 mag \citep{Schlafly2011}. The uncertainties are dominated by the zero point uncertainties ($\sim0.05$\,mag). The resulting colour of $(g - r)_0 = 0.54 \pm 0.08 $ mag is typical for old stellar populations dominated by RGB stars such as in early-type dwarf galaxies (e.g. \citealt{Mueller2018a}). To compare the NSC to a larger sample of literature sources, we converted these magnitudes into Johnson $V$-band using the relation from Lupton 2005\footnote{\url{https://www.sdss3.org/dr10/algorithms/sdssUBVRITransform.php} of $V = g - 0.5784 (g-r) - 0.0038$ and a distance modulus of 27.63 \citep{Mueller2019}. We obtained an absolute magnitude of $M_V$ = $-$9.51 $\pm$ 0.07 mag.}

\begin{figure}
    \centering
    \includegraphics[width=0.49\textwidth]{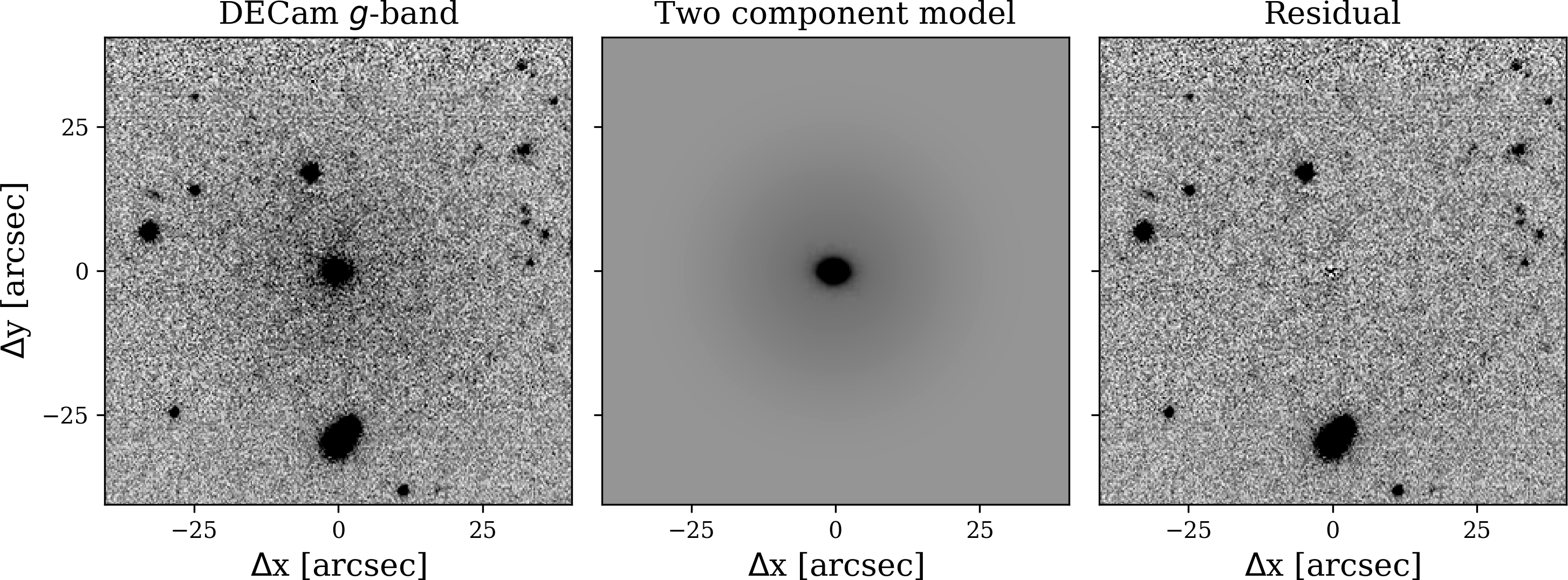}
    \caption{Illustration of the two component \textsc{imfit} model. Left: Original DECam $g$-band image of KKs58. Middle: double S\'{e}rsic model. Right: residual after subtracting the model from the image. North is up, east is to the left.}
    \label{fig:imfit_model}
\end{figure}

We used the two-dimensional image fitting routine \textsc{imfit} \citep{Erwin2015} to determine the structural parameters of the NSC. Our best-fit model, shown in Fig. \ref{fig:imfit_model}, consists of two S\'{e}rsic models, one representing the NSC and the second corresponding to the underlying galaxy. This model considers a model of the PSF that we built with the effective PSF functionality of the python package \textsc{photutils} \citep{Bradley2019} using $\sim$ 10 stars in the FOV, following the prescription of \citet{Anderson2000}. We found effective radii of the NSC and the host of 6.6 $\pm$ 0.5 pc (0.40 $\pm$ 0.02\arcsec) and \mbox{244.7 $\pm$ 8.3 pc} (15.0 $\pm$ 0.5\arcsec), respectively. The uncertainties were inferred with Markov-Chain Monte Carlo (MCMC) sampling. The effective radius of the host galaxy is lower than earlier measurements of 430 pc from \cite{Jerjen2000}, which could be due to better image quality and depth of the DECam images.
According to our model, the NSC is elongated with an ellipticity of 0.30 $\pm$ 0.04 at a position angle of 86 $\pm$ 2 $^\circ$ clockwise from North, while the galaxy S\'{e}rsic model is completely spherical. 

\subsection{KK197 HST ACS photometry}
KK197 was originally studied by \cite{Georgiev2009a} using the F606W and F814W filters of the HST ACS instrument with the aim of identifying GC candidates in nearby dwarf galaxies. The observations were obtained in HST cycles 12 and 13 (PI: I. Karachentsev), and are described in \cite{Georgiev2008} and \cite{Georgiev2009a}. The identification of GC candidates was based on colour constraints and their (resolved) morphologies. For KK197, \cite{Georgiev2009b} reported three GC candidates and we list their properties in Tab. \ref{tab:NSCs_and_host_properties}. The largest, dubbed KK197-02, is located at the photometric centre of KK197 and in the following we refer to this source as KK197-NSC. KK197-NSC has a similar brightness than KKs58-NSC, but is slightly smaller. KK197-01 and KK197-03 turned out to be indeed GCs, as our MUSE analysis confirmed (Sect. \ref{sect:KK197_MUSE_GCs}). 

%------------------------------------MUSE spectroscopy ------------------------------------------------------------
\section{MUSE spectroscopic analysis}
\label{sect:spect}
In the following we present our analysis of the MUSE data of KKs58 and KK197. In particular, we distinguish between the star clusters and the host galaxies. The same method of full spectral fitting, described in the next section, is applied to both. 
\begin{figure*}
\centering
\includegraphics[width=0.99\textwidth]{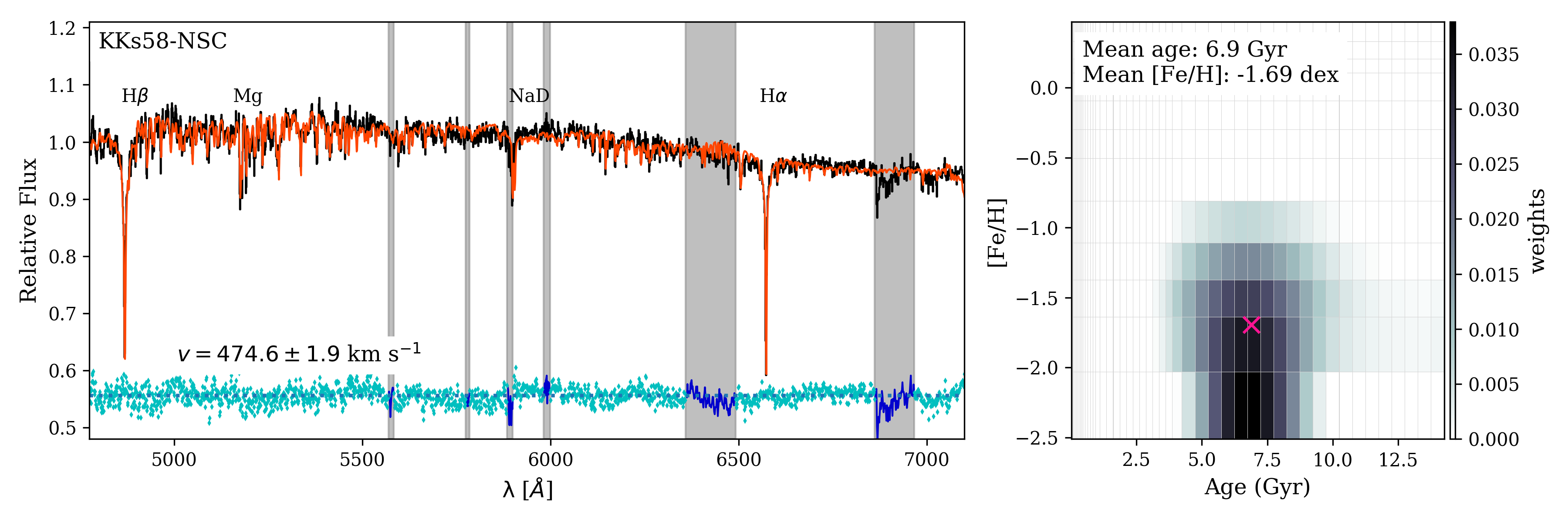}
\includegraphics[width=0.99\textwidth]{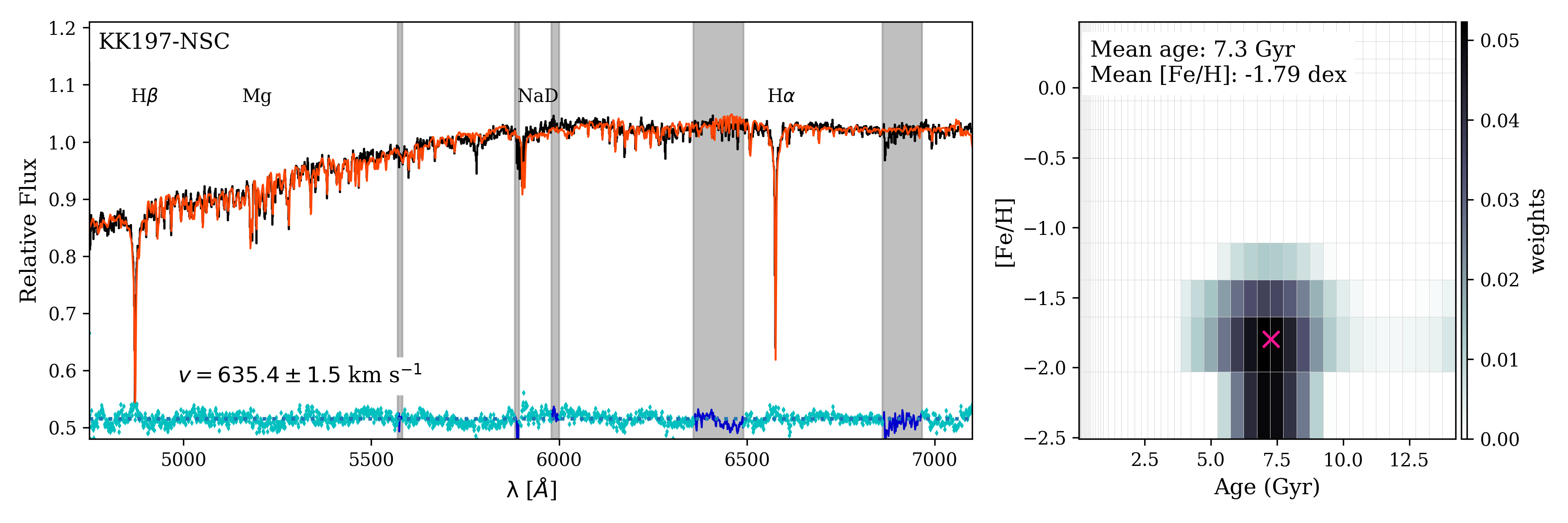}
\caption{Normalised spectrum of KKs58-NSC (top) and KK197-NSC (bottom) fitted with \textsc{pPXF} and the scaled-solar MILES models. The input spectrum of the NSC is shown in black and the best-fitting combination of MILES SSP models is shown in red. The blue points indicate the residual, shifted to 0.5 for visibility. Masked regions with strong sky residual lines appear as grey-shaded. The right panels show the available grid of MILES SSP models in age and metallicity, colour-coded by their weight in the best-fit. The pink crosses mark the position of the weighted mean ages and metallicities. These are not corresponding to our final measurements from MC fitting listed in Tab. \ref{tab:NSCs_and_host_properties}.}
\label{fig:NSC_spec_fit}
\end{figure*}

\begin{figure*}
\centering
\includegraphics[width=0.30\textwidth]{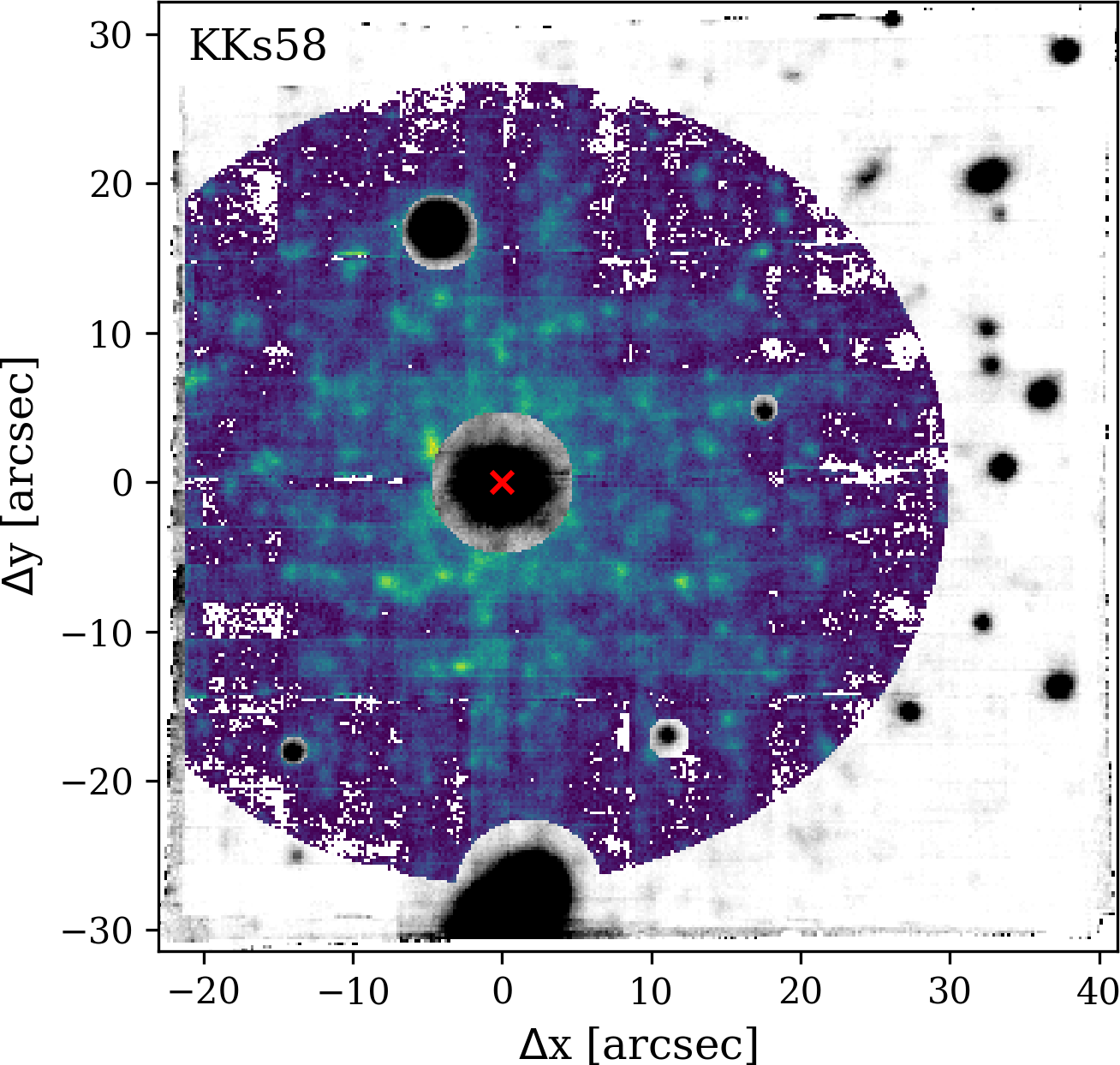}
\includegraphics[width=0.69\textwidth]{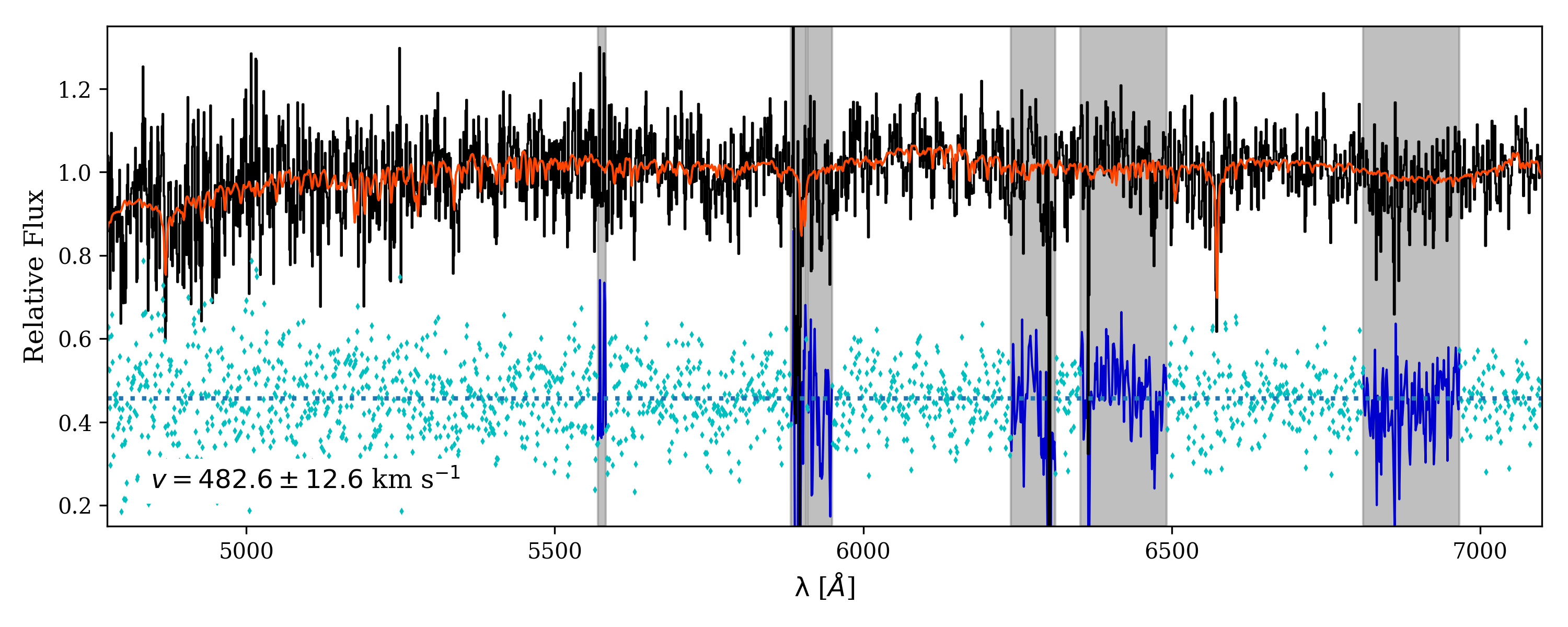}
\caption{Extraction of the KKs58 galaxy spectrum. Left: White-light image of KKs58, created by collapsing the MUSE cube along the spectral axis (10\arcsec $\approx 160$ pc). The position of the NSC is marked by the red cross. We show the mask that is used for the extraction of the integrated galaxy spectrum in green-blue colours while the masked pixels are shown with are in black and white. Right: \textsc{pPXF} fit to the normalised integrated spectrum of KKs58. The input spectrum is shown in black and the best-fitting combination of MILES SSP models is shown in red. The blue points indicate the residual, shifted to 0.5 for visibility. Masked regions with strong sky residual lines appear as grey-shaded. We only use the spectral range from 4500 to 7100 \AA\,because the spectrum is heavily contaminated by sky residual and telluric lines at larger wavelengths.}
\label{fig:KKs58_gal_map}
\end{figure*}

\subsection{Full spectral fitting of MUSE spectra}
\label{sect:ppxf}
We fitted the MUSE spectra using the penalised Pixel-fitting routine (\textsc{pPXF}, \citealt{Cappellari2004, Cappellari2017}). \textsc{pPXF} is a method for full spectral fitting that uses a penalised maximum likelihood approach to fit a linear combination of template spectra to a given input spectrum. This enables extraction of the line-of-sight (LOS) velocity distribution, but also to determine stellar population properties. We used the single stellar population (SSP) template spectra from the Medium resolution INT Library of Empirical Spectra (MILES, \citealt{Vazdekis2010}). The MILES templates give the spectral energy distribution in a wavelength range from 3525 to 7500 \AA\,for stellar populations of a single age and metallicity. We used BaSTI isochrones \citep{Pietrinferni2004, Pietrinferni2006} and a Milky Way-like, double power law (bimodal) initial mass function (IMF) with a high-mass slope of 1.30 \citep{Vazdekis1996}. Generally, the MILES models are classified by their age and total metallicity [M/H]. However, for comparisons to photometric measurements, we used the scaled-solar MILES models that do not include $\alpha$-enhancement and thus have [M/H] = [Fe/H]. The MILES spectra have a spectral resolution of 2.5 \AA\ \citep{FalconBarroso2011}, same as the mean instrumental resolution of MUSE.

Throughout this work, we fitted LOS velocities using additive polynomials of degree 12 and no multiplicative polynomials. When fitting for stellar population properties, we keep the LOS velocity fixed and used multiplicative polynomials of degree 8 without any additive polynomials. In all cases, the velocity dispersions of the star clusters and host galaxies are close or below the MUSE resolution and are not accessible. To obtain realistic uncertainties on the LOS velocity, the mean metallicity and age, we used a Monte Carlo (MC) approach (e.g. \citealt{Cappellari2004, Pinna2019}). After the first fit, we drew random values from the \textsc{pPXF} residual in every wavelength bin and add these values to the first best-fit spectrum. This way, a new realisation of the spectrum is created that can be fitted again. This procedure is repeated 100 times to obtain a well sampled distribution around the mean LOS velocity, the metallicity and the age. The uncertainties are then given by the standard deviation of the resulting distribution. 

In general, we distinguish the spectra by their spectral signal-to-noise ratio (S/N), obtained in a continuum region around 6000 \AA. In the worst cases, for S/N $\leq$ 3 \AA$^{-1}$, no reliable information can be drawn from the spectrum. For S/N $>$ 3 \AA$^{-1}$, at least the line-of-sight (LOS) velocity is measurable to confirm membership of a star cluster to its host. Fitting spectra with \mbox{S/N $\geq$ 10 \AA$^{-1}$} further gives reliable estimates of the mean metallicity. In the following, we give the ages as obtained from MC fitting with \textsc{pPXF} with their formal errors that are typically on the order of 1 Gyr, but these uncertainties might underestimate the true uncertainties due to the lack of strong age-sensitive features in the MUSE spectrum (see the Appendix in \citealt{Fahrion2019b}). In addition, even when provided with a prior from photometry, measuring accurate ages of old stellar populations with integrated spectra is challenging \citep{Usher2019} and the resulting ages can deviate from the true ages by up to 5 Gyr. Age estimates of young populations $<$ 5 Gyr, however, have a higher accuracy.
For spectra with S/N $>$ 50 \AA$^{-1}$, such as the spectra of the NSCs, full spectral fitting with regularisation allows studying star formation histories and possible multiple populations. The regularisation ensures a smooth distribution of weights, that is needed, for example, to extract star formation histories \citep{Boecker2019}. Because the MILES models are normalised to 1 $M_\sun$, our stellar population properties are mass-weighted. 

Due to non-photometric observing conditions, the flux calibration of our MUSE spectra has substantial systematic errors visible as different continuum shapes in KKs58 and KK197 (see Fig. \ref{fig:NSC_spec_fit}). However, this should not affect our results because of the polynomials that are used by \textsc{pPXF} to account for the continuum variation that are not taken into account as constraint for stellar population determination.

\subsection{KKs58 nuclear star cluster}
\label{sect:NSC_spec}
The NSC of KKs58 is clearly visible in the MUSE FOV (Fig. \ref{fig:KKs58_gal_map}). The NSC appears to be slightly elongated, similar to what we found in the FORS2 and DECam data. However, to extract its spectrum, we treated the NSC as a point source and used a circular aperture weighted by the PSF assuming a Gaussian profile with FWHM = 1.0\arcsec. This weighting helps to boost the signal of the NSC compared to the faint underlying galaxy background.

The NSC spectrum has S/N $\sim$ 85\,\AA$^{-1}$. From 100 MC fits to the spectrum, we found a heliocentric LOS velocity of KKs58's NSC of \mbox{v$_{\text{KKs58-NSC}} = 474.6\,\pm\,1.9\,\text{km s}^{-1}$}. The top panel of Fig. \ref{fig:NSC_spec_fit} shows the original NSC spectrum and the regularised \textsc{pPXF} fit with a regularisation parameter of 70.
We found a mass-weighted mean metallicity of $[$Fe/H$]$ = $-1.69$ dex and a weighted mean age of $\sim$ 7 Gyr. Testing with other stellar libraries highlights how challenging the determination of old ages is, because if we used the extended MILES library for the fitting, we found a mean age of $\sim$ 9 Gyr. Nevertheless, the \textsc{pPXF} fit to the NSC spectrum can exclude recent star formation ($<$ 2 Gyr) in the NSC. The regularised fit does not show multiple populations of different ages and metallicities. However, with the used SSP template grid and spectral quality, we are unable to detect the small metallicity variations of $\sim$ 0.1 dex that can be found in massive MW GCs such as $\omega$Cen and M54 \citep{Marino2015, Johnson2015}. 
Therefore, we use 100 MC fits without regularisation to determine reliable measurements of mean age and metallicity. We derived an age of 6.9 $\pm$ 1.0 Gyr and $[$Fe/H$]$ = $-$1.75 $\pm$ 0.06 dex. The given uncertainties refer to random errors, but the systematic uncertainties can be larger. Both are consistent with the weighted means from the regularised fit.

We can use stellar population analysis of the NSC to estimate its stellar mass using the photometric predictions for the mass-to-light ratio ($M/L$) for a given SSP of the scaled-solar MILES models. We used the predictions for the Johnson $V$-band and determined the total luminosity using a randomly drawn $V$-band magnitude assuming a Gaussian distribution of the magnitude with \mbox{$M_V = -9.51$} mag and $\sigma = 0.07$ mag. This was repeated 5000 times to accommodate for the magnitude uncertainties. We found a total luminosity of \mbox{$L_{V,\, \text{KK197-NSC}}$ = 5.36 $\pm$ 0.35 $\times 10^5 L_\sun$} and \mbox{$M/L_V$ = 1.37 $\pm$ 0.15 $M_\sun/L_\sun$}, corresponding to the expected $M/L$ from our MC \textsc{pPXF} fit. This gives a stellar mass of \mbox{$M_{\ast, \text{KK197-NSC}} = 7.34 \pm 0.87 \times 10^5 M_\sun$}. KKs58-NSC is therefore less massive than the two most massive GCs in the MW, $\omega$\,Cen and M\,54 that have masses of \mbox{3.6 $\times 10^6 M_\sun$} and \mbox{$1.4 \times 10^6 M_\sun$}, respectively \citep{BaumgardtHilker2018}.

\subsection{KKs58 galaxy spectrum}
\label{sect:KKs58_gal_spec}
The galaxy KKs58 itself is very faint in the MUSE data as Fig. \ref{fig:KKs58_gal_map} illustrates. We attempted to bin the MUSE cube with the Voronoi binning scheme \citep{CappellariCopin2003} to obtain a binned map of KKs58, but this was unsuccessful due to the low S/N. So we only extracted a single spectrum of the galaxy (right panel in Fig. \ref{fig:KKs58_gal_map}) using a mask applied to the full cube (Fig. \ref{fig:KKs58_gal_map} left). 
Based on the galaxy shape in the DECam data and deep FORS2 image, this mask is circular with a radius of 120 pixels (24\arcsec $\approx$ 390 pc), centred on the NSC, and excludes spaxels that have a negative mean flux, the NSC, and several foreground stars.

The final galaxy spectrum has a S/N of 11 \AA$^{-1}$ and is heavily contaminated by sky residual and telluric lines above $\sim$ 7500\,\AA. The \textsc{pPXF} fit to the galaxy spectrum is shown in Fig. \ref{fig:KKs58_gal_map}. Unfortunately, the S/N of the galaxy spectrum is not sufficient to explore the stellar population properties as we did it for KKs58-NSC. Nonetheless, we determined the LOS velocity, mean age and metallicity with 100 MC fits as described above. We measured a mean LOS velocity for KKs58 of \mbox{v$_\text{KKs58} = 482.6\,\pm\,12.6\,\text{km s}^{-1}$}, fully consistent with the LOS velocity of the NSC. We obtained a mean metallicity of [Fe/H] = $-$1.35 $\pm$ 0.23 dex and an age of $\sim$ 7 Gyr.
\cite{Mueller2019} give a photometric metallicity estimate taken from fitting the FORS2 colour magnitude diagram with old (10~Gyr) isochrones of \mbox{[Fe/H] = $-$1.49 $\pm$ 0.80 dex}, in agreement with our spectroscopic measurement. 

We determined the stellar mass of KKs58 using the stellar population properties as obtained from the \textsc{pPXF} fit, deriving a $M/L_V$ of 1.46 $\pm$ 0.15 $M_\sun/L_\sun$. With a total luminosity of \mbox{$L_{V, \text{KKs58}}$ = 5.02$^{+0.55}_{-0.05} \times 10^6 L_\sun$ \citep{Mueller2019}}, this translates to a total stellar mass of $M_{\ast, \text{KKs58}} = 7.33^{+1.10}_{-0.76} \times 10^6 M_\sun$. Therefore, the NSC-to-galaxy mass ratio is $\sim 10\%$, a typical value for a galaxy with such a mass (e.g. \citealt{SanchezJanssen2019}).

%%%%%%%%%------------------ KK197 NSC + GCs

\begin{figure}
    \centering
    \includegraphics[width=0.49\textwidth]{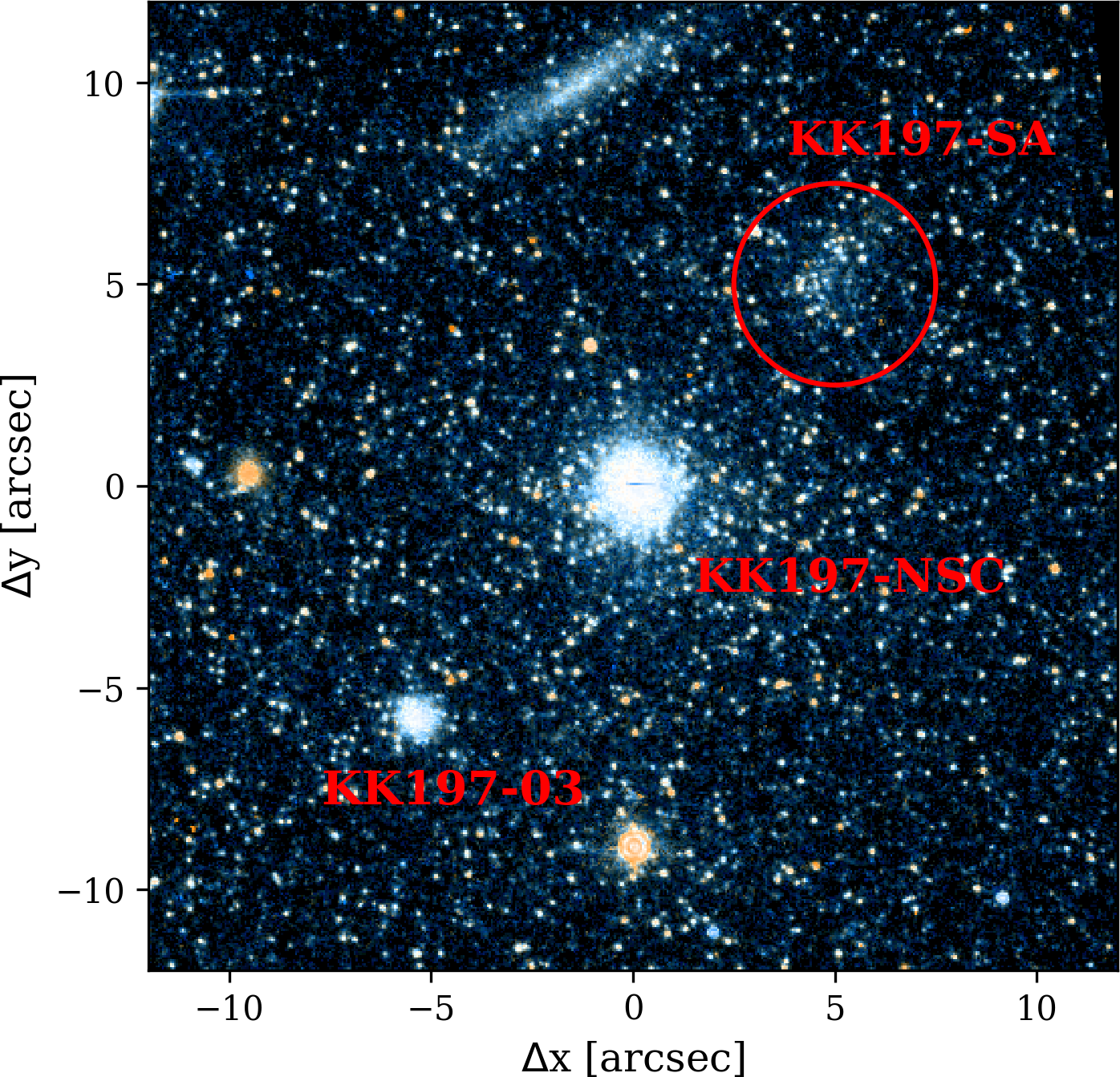}
    \caption{HST colour image of the central part of KK197, using the F606W and F814W filters (10\arcsec $\approx$ 190 pc). The red circle shows the loose stellar association (KK197-SA).}
    \label{fig:KK197_blob}
\end{figure}

\begin{figure*}
\includegraphics[width=0.99\textwidth]{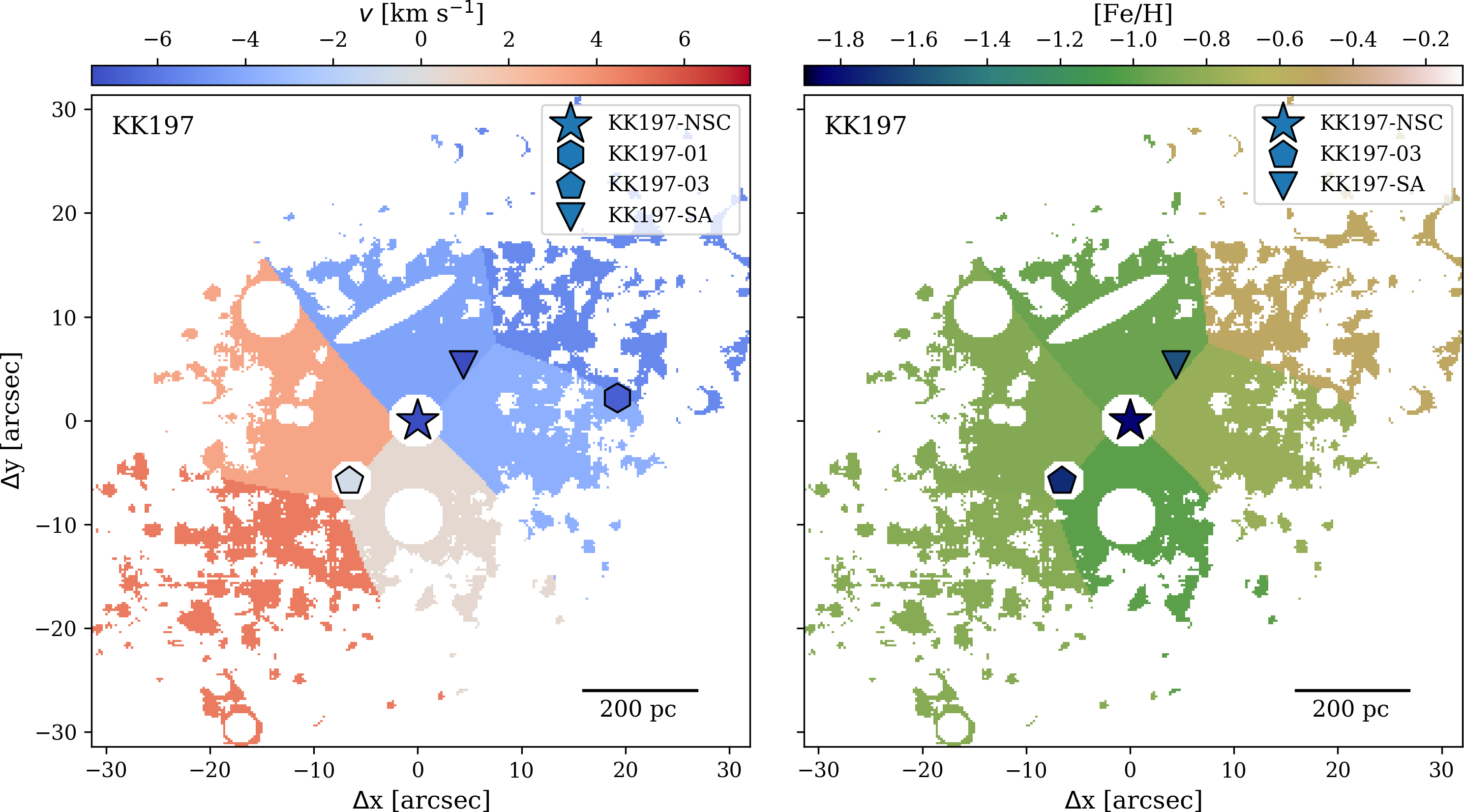}
\caption{Voronoi binned maps of KK197 in comparison to the star clusters. The maps were binned to S/N = 25 \AA$^{-1}$. Left: LOS velocity map relative to the median velocity of 643.2 km s$^{-1}$. We added the LOS velocities of KK197-NSC, the two GCs and the loose stellar association of stars (KK197-SA) described in Sect. \ref{sect:blob}. Right: Mean metallicity map of KK197 obtained from an unregularised \textsc{pPXF} fit using the scaled-solar MILES SSP models.}
\label{fig:KK197_maps}
\end{figure*}

\begin{figure}
    \centering
    \includegraphics[width=0.49\textwidth]{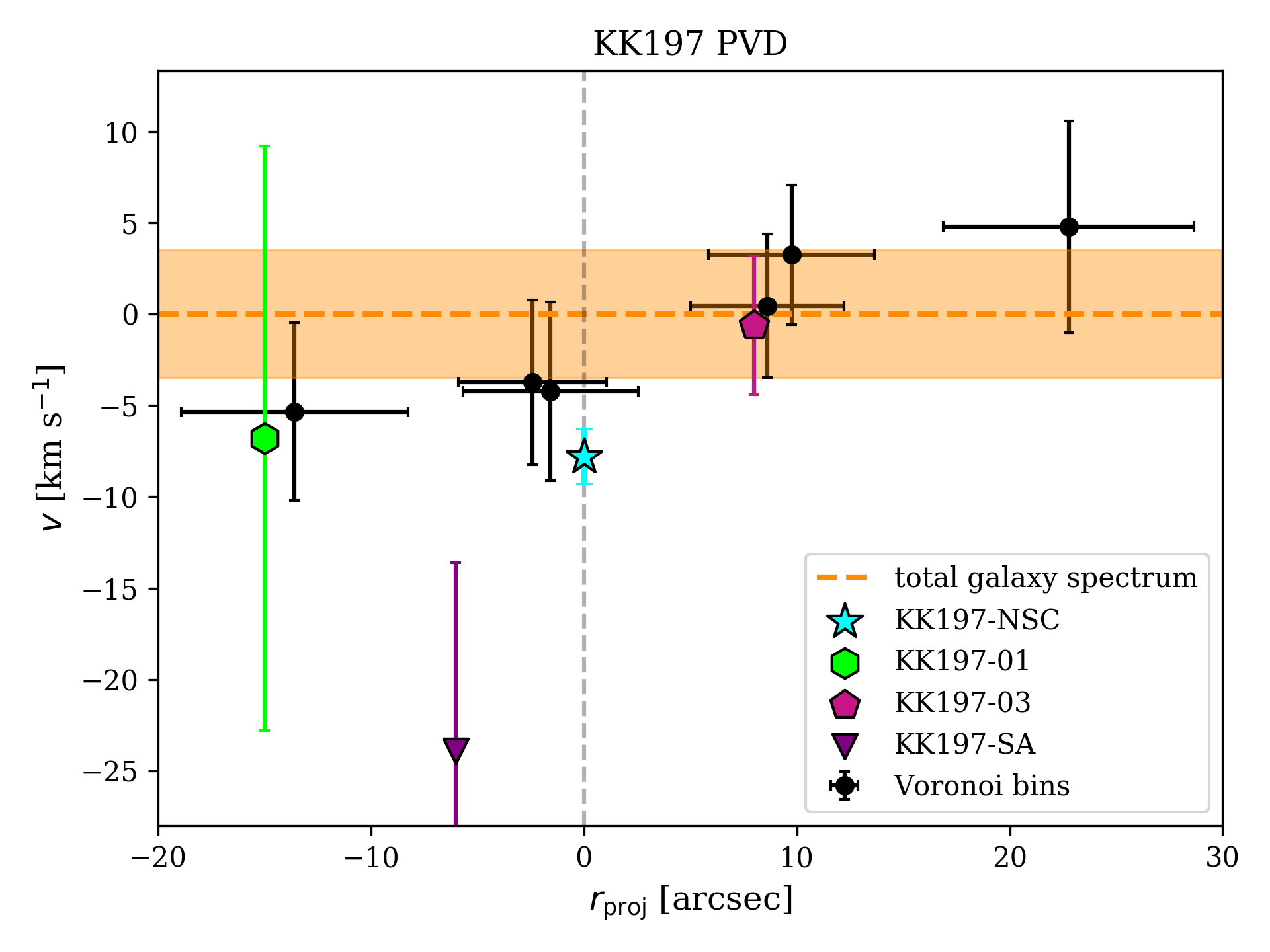}
    \caption{Position-velocity diagram for KK197. The NSC, the GCs and the stellar association (KK197-SA) are marked by the coloured symbols. The x-axis is the distance projected on the major axis of the galaxy. The errorbars refer to the uncertainties from 100 MC fits. The orange coloured region shows the velocity obtained from the total galaxy spectrum of 463.2 $\pm$ 3.5 km s$^{-1}$.}
    \label{fig:KK197_PVD}
\end{figure}

\subsection{KK197 nuclear and globular star clusters}
\label{sect:KK197_MUSE_GCs}
We extracted the spectra of KK197-NSC and the two GC candidates KK197-01 and 03 using a PSF-weighted circular aperture with a FWHM of 1.0\arcsec and fitted these spectra with \textsc{pPXF}. The results of these fits can be found in Tab. \ref{tab:NSCs_and_host_properties}. For the NSC of KK197, we found a LOS velocity of v$_{\text{KK197-NSC}} = 635.4 \pm 1.5$ km s$^{-1}$. 

KK197-NSC has a spectrum with a S/N of $\sim$ 100\,\AA$^{-1}$, sufficient to analyse the stellar populations properties such as age and metallicity distributions using a regularised fit (see bottom panel in Fig. \ref{fig:NSC_spec_fit}). The weighted mean age and metallicity from the regularised fit are 7.3 Gyr and $-$1.79 dex, respectively, similar to the results for the NSC of KKs58. Again, the regularised fit gives no indication of multiple populations with the used SSP grid. Smaller scale variations such as small metallicity spreads as observed in some MW GCs cannot be ruled out. The 100 MC fits without regularisation give an age of 6.5 $\pm$ 1.0 Gyr and \mbox{[Fe/H] = $-1.84 \pm 0.05$ dex}, in agreement with the weighted means of the regularised fit.
The bright GC KK197-03 has a spectral S/N of 31\,\AA$^{-1}$, still sufficient to analyse the stellar populations. With 100 MC fits, we found \mbox{[Fe/H] = $-$1.80 $\pm$ 0.11 dex} and the LOS velocity of \mbox{v$_\text{KK193-03} = 642.6 \pm 3.8$ km s$^{-1}$} clearly confirms membership to KK197. Our fit resulted in a mean age of $\sim$ 7 $\pm$ 1 Gyr in agreement with the NSC, but at this S/N the age is uncertain.

We determined the stellar masses of KK197-NSC and KK197-03 using the photometric predictions for the scaled-solar MILES models in the $V$-band. For KK197-NSC, we found $L_{V, \text{KK197-NSC}} = 7.60 \pm 0.42 \times 10^5 L_\sun$ and $M/L_V = 1.37 \pm 0.13\,M_\sun/L_\sun$. This results in a stellar mass of $M_{\ast, \text{KK197-NSC}}= 1.04 \pm 0.11 \times 10^6 M_\sun$. The GC KK197-03 has a total luminosity of $L_{V, \text{KK197-03}} = 7.12 \pm 0.39 \times 10^4 L_\sun$ and with a $M/L_V$ = 1.37 $\pm$ 0.14 $M_\sun/L_\sun$, we obtained a mass of $M_{\ast, \text{KK197-03}} = 9.76\,\pm\,1.13\,\times\,10^4\,M_\sun$. KK197-01, the faintest star cluster in KK197 identified by \cite{Georgiev2009a}, has as S/N of 8 \AA$^{-1}$, insufficient to obtain a reliable metallicity estimate. We can confirm its membership to KK197 based on its radial velocity of \mbox{636.4 $\pm$ 16.0 km s$^{-1}$}. 

By inspecting other point sources in the MUSE FOV, we identify two other potential faint star clusters (RA = 13:22:03.0, DEC = $-$42:32:07.5 and RA = 13:22:01.2, DEC = $-$42:32:18.3). Their MUSE spectra have low S/N of 6.2 and 4.5 \AA$^{-1}$, respectively. Their measured LOS velocities are in agreement with KK197 of v$ = 652 \pm 31$ km s$^{-1}$ and v$ = 663 \pm 69$ km s$^{-1}$, respectively. However, inspection of the HST image suggests that those are stars of KK197 blended along the line of sight.

\subsection{Dynamical mass estimate of KK197-NSC}
The NSC of KK197 was observed on 11 January 2010 with the VLT's UVES spectrograph \citep{Dekker2000} as part of the programme 084.D-0818 (PI: I. Georgiev). The instrumental setup included the red spectrograph arm with the standard setting centered on 580 nm, $2 \times 2$ CCD binning and a $1\farcs2$ slit width, which yielded a spectral resolution of R=34540, suitable for accurate velocity dispersion measurement. A single exposure of 3004 seconds resulted in a S/N ratio of 7 -- 11 over the spectral range  4786 -- 5761\,\AA\, covered with the lower CCD and S/N of 10--15 over 5835 -- 6808\,\AA\, covered with the upper CCD. We downloaded the raw spectrum  with the associated calibrations from the ESO Science Archive and reduced it using the ESO UVES pipeline (version 5.10.4) within the Reflex environment \citep{Freudling2013}. The velocity dispersion was measured by cross-correlating the spectrum of the KK197 NSC with a set of 18 UVES spectra of 13 different G and K-type giant stars having a range of metallicities  ($-2.6<$ [Fe/H] $<+0.3$~dex). These stars were observed as part of the UVES programme 069.D-0169 (PI: Rejkuba), which used the same red arm and 580 nm spectral setting, but had $1\farcs 0$ slit width and no detector binning. We adopted the same procedure to measure the velocity dispersion as described in \citet{Rejkuba2007}. First, given the difference in slit width and detector binning setup between the KK197-NSC and the giant star observations,
we verified the dependency of velocity dispersion measurements on metallicity and instrumental broadening both with empirical templates and with synthetic spectra of K5 giants that were computed with different resolutions using the online spectrum synthesis tool developed and maintained at MPIA \citep{NLTE_MPIA}. The Fourier-filtered spectrum of the NSC was cross-correlated with all template spectra within the IRAF fxcor task. The highest cross-correlation signal was measured for the stars with similar metallicity, and therefore for the final result we averaged the measurements for which the fxcor \textit{TDR} index was $>22$ \citep[see][]{tonry1979}. This resulted in the final velocity dispersion of $25.2\pm 0.6$~km s$^{-1}$ for the lower and $24.1 \pm 2.2$ km s$^{-1}$ for the upper CCD. The median velocity of the NSC obtained from cross-correlation with template stars is $636.68 \pm 0.77$~km s$^{-1}$, fully consistent with the velocity measured from the MUSE data.

In order to derive a dynamical mass for KK197-NSC, an aperture correction has to be applied to convert the observed velocity dispersion in the extracted slit area to a global and central velocity dispersion. We follow the same approach as described in \citet{Hilker2007}. In short, we first constructed a 3-dimensional density profile from the structural parameters of a King profile fit \citep{King1962} to HST data (projected half-light radius $r_h=3.03$\,pc at a distance of 3.98\,Mpc (this paper) and concentration $c= 1.48$, \citealt{Georgiev2009a}). From that the cumulative mass distribution $M(<r)$, the potential energy $\phi(r)$, and the energy distribution function $f(E)$ are calculated. The corresponding 6D phase space density distribution is then sampled by a N-body representation of the NSC with $10^5$ particles. $x$, $y$, $z$ positions and $vx$, $vy$ and $vz$ velocities are assigned to each particle, assuming spherical symmetry and an isotropic velocity dispersion. The influence of seeing is modelled by assuming that, in a projected version of the N-body model, the light from each particle is distributed as a 2D Gaussian whose FWHM corresponds to the observed seeing of $1\farcs 0$. For each particle, the fraction of the light that falls within the spectroscopic extraction aperture of $1\farcs 2\times 4\farcs 0$ is calculated. The fraction of light in that aperture is used as weighting factor for the velocities. The weighted velocity contributions are then used to calculate the expected velocity dispersion in the extracted slit area. Finally, the mass of the N-body model is iteratively adjusted such that the calculated velocity dispersion matches the observed one.

The model dynamical mass that matches the observed velocity dispersion of $25.2\pm 0.6$~km s$^{-1}$ for the lower CCD is \mbox{$M_{\rm KK197-NSC,\,dyn}=4.3\pm0.3 \times 10^6 M_\sun$}. The global velocity dispersion of the NSC is 24.8\,km s$^{-1}$ and the central one 30.4\,km s$^{-1}$, according to the modelled parameters.
Our derived dynamical mass is about four times higher than the stellar mass derived from stellar population models. This difference might be caused by systematic uncertainties in the stellar population properties of the NSC and the assumptions of our mass modelling. We further discuss possible caveats and the implications of that difference in Sect. \ref{sect:UCD_dis}.

\subsection{Stellar association in KK197}
\label{sect:blob}
North-east of the KK197-NSC is an association of stars (see Fig. \ref{fig:KK197_blob}) that has a higher number density than the surrounding galaxy body and has a diameter of $\sim$ 33 pc (1.75\arcsec) in the HST image. We refer to this stellar association as KK197-SA. We extracted the spectrum of this region by selecting the pixels belonging to it above a flux threshold. The spectrum has a S/N of 11.1 \AA$^{-1}$ and fitting for the velocity and mean metallicity with 100 MC fits, we found v$_\text{KK197-SA} = 619.3 \pm 10.3$ km s$^{-1}$ and \mbox{[Fe/H] = $-1.59 \pm 0.15$ dex}. With the low S/N, we cannot determine the age robustly, but our fits give a mean age of $\sim$ 10 Gyr, similar to the host galaxy. We cannot see any emission lines in the spectrum. This association of stars is clearly connected with KK197 and although the associated random errors are quite large, it appears to be more metal-poor than the host galaxy (mean metallicity of $-0.84 \pm 0.12$ dex, see Fig. \ref{fig:KK197_maps}). The morphology excludes a classical GC. This stellar association might resemble an open star cluster, or maybe a GC in the process of being disrupted in the central potential of KK197.

\begin{table*}
    \centering
    \caption{Properties of KKs58, KK197, their NSCs and GCs. 
The quoted errors on v$_\text{LOS}$, [Fe/H] and age were derived with 100 fits to the respective spectra.}
    \begin{threeparttable}
    \begin{tabular}{ c  c  c  c  c  c  c  c  c}\hline \hline
    Object & RA & DEC & $M_V$ & r$_\text{eff}$ & S/N & v$_\text{LOS}$ & $[$Fe/H$]$  & M$_\ast$ \\
            & (J2000) & (J2000) & (mag) & (pc) & (\AA$^{-1}$) & (km s$^{-1}$) & (dex)  & (10$^5 M_\sun$) \\\hline
    KKs58   & 13:46:00.8 & $-$36:19:44 & $-11.93^{+0.12}_{-0.01}$\tnote{a} & 244.7 $\pm$ 8.3 & 11.1 & 482.6 $\pm$ 12.6 & $-1.35 \pm 0.23$ &  $73.3^{+11}_{-7.6}$ \\
    KKs58-NSC & 13:46:00.8 & $-$36:19:44 & $-$9.51 $\pm$ 0.07 & 6.6 $\pm$ 0.5 & 85.3 & 474.6 $\pm$ 1.9 & $-$1.75 $\pm$ 0.06  & 7.34 $\pm$ 0.87 \\
    KK197 & 13:22:02.0 & $-$42:32:08.1 & $-13.04$\tnote{b} & 733\tnote{c} & 21.8 & 643.2 $\pm$ 3.5 & $-$0.84 $\pm$ 0.12 & $\sim$ 400 \\
    KK197-NSC & 13:22:02.0 & $-$42:32:08.1 & $-$9.89 $\pm$ 0.06\tnote{b} & 3.03 $\pm$ 0.13\tnote{b} & 99.8& 635.4 $\pm$ 1.5 & $-$1.84 $\pm$ 0.05 & 10.4 $\pm$ 1.1 \\
    KK197-01 & 13:21:59.8 & $-$42:32:06.5 & $-$5.75 $\pm$ 0.07\tnote{b} & 2.01 $\pm$ 0.17\tnote{b} & 8.1 & 636.4 $\pm$ 16.0 & -- & -- \\
    KK197-03 & 13:22:02.5 & $-$42:32:13.8 &  $-$7.32 $\pm$ 0.06\tnote{b} & 2.63 $\pm$ 0.17\tnote{b} & 31.0 &  642.6 $\pm$ 3.8 & $-$1.80 $\pm$ 0.11 &  0.98 $\pm$ 0.11  \\
    KK197-SA & 13:22:01.6 & $-$42:32:02.6 & -- & $\sim 16$ & 11.1 & 619.3 $\pm 10.3$ & $-1.59 \pm 0.15$ & -- \\
    
 \hline
\end{tabular}
\begin{tablenotes}
\item[a] from \cite{Mueller2019}
\item[b] from \cite{Georgiev2009a}, updated to a distance of $D = 3.87$ Mpc.
\item[c] from \cite{Sharina2008}
\end{tablenotes}
    \label{tab:NSCs_and_host_properties}
    \end{threeparttable}
\end{table*}

\subsection{Galaxy KK197}
\label{sect:KK197_MUSE_gal}
The MUSE data of KK197 has a sufficient S/N to obtain a binned map of the integrated light after applying an ellipse cutout around the NSC. We masked bright stars, background galaxies, and the known star clusters in the FOV. The MUSE data of KK197 was binned to a continuous S/N of 25 \AA$^{-1}$ using the Voronoi-binning scheme described in \cite{CappellariCopin2003}. This map contains six bins. The binned spectra were fitted with \textsc{pPXF} with 100 MC runs each to acquire maps of the LOS velocities and the metallicities. Ages are uncertain at this S/N level, but we show the LOS velocity map and the mean metallicity map in Fig. \ref{fig:KK197_maps}. In these maps, we overplotted the properties of the NSCs and the GCs for comparison with the stellar light.
In addition, Fig. \ref{fig:KK197_PVD} shows the position-velocity diagram for KK197. For this, we projected the bins onto the major axis and show their LOS velocity relative to the velocity obtained from the total galaxy spectrum of \mbox{643.2 km s$^{-1}$}.

The LOS velocity field shows a rotation signal along the major axis of the galaxy with a maximum amplitude of $\sim$ $\pm$ 5 km s$^{-1}$. However, the Voronoi bins have typical LOS velocity uncertainties of $\sim$ 4.5 km s$^{-1}$. We tested different binning schemes and obtained the same rotation signal robustly throughout these tests. The star clusters we identified in KK197 seem to be in agreement with the rotation signature, but deeper data is required to confirm it. Interestingly, the NSC might not be at rest with respect to the galaxy. Instead, from the total galaxy spectrum obtained from all six bins, we found a LOS velocity of \mbox{v$_{\text{KK197}} = 643.2 \pm 3.5$ km s$^{-1}$}, corresponding to a velocity difference between NSC and host of \mbox{$\Delta v = 7.8 \pm 3.8$ km s$^{-1}$}. A higher S/N in the galaxy spectrum would be required to confirm this offset. 

KK197-NSC appears as a distinct component in the metallicity map shown in the right panel of Fig. \ref{fig:KK197_maps}. It is significantly more metal-poor than the galaxy field star population, for which we obtain a mean metallicity of \mbox{[Fe/H] = $-0.84\,\pm$ 0.12 dex}. This is an intriguing result because in massive galaxies an opposite trend is found and the central region show higher metallicities due to efficient star formation fuelled by infalling, enriched gas (e.g. \citealt{Spengler2017}). Our metallicity estimate of the host galaxy is in agreement with the mean photometric metallicity of \mbox{[Fe/H] = $-$1.05 dex} determined by \cite{Crnojevic2010} using the HST/ACS data.
From the combined spectrum of KK197, we obtained a mean age of 10 $\pm$ 1 Gyr. 

Using the fit to the total spectrum of KK197 and its $V$-band magnitude of $-13.04$ mag \citep{Georgiev2008}, we derived a total stellar mass of \mbox{$M_{\ast, \text{KK197}} \sim 4 \times 10^7 M_\sun$}. Based on this rough mass estimate, the ratio of NSC-to-host mass is $\sim$3 \%, smaller than what is observed in KKs58. 

% ------------- Discussion
\section{Discussion}
\label{sect:discussion}
In the following, we place our measurements
for KKs58 and KK197 into context with respect to other nucleated dwarf galaxies. We further discuss how the properties of the NSCs constrain their formation pathway and how the NSCs compare to UCDs. 

\subsection{Comparison to other nucleated galaxies}
Both in KKs58 and KK197, we found metal-poor NSCs with metallicities comparable to metal-poor GCs. This is in agreement with a study of 61 NSCs of dwarf galaxies in the Fornax galaxy by \cite{Ordenes2018} that found that most most dwarf nuclei have colours consistent with metal-poor GCs. They also report a bimodal mass distribution of dwarf galaxy NSCs with peaks at log($M_\ast/M_\sun) \approx 5.4$ and 6.3. With masses of log($M_\ast/M_\sun) = 5.9$ and 6.0, respectively, the NSCs of KKs58 and KK197 are located in between the two mass peaks.

We estimated the NSC-to-host stellar mass of KKs58 and KK197 to be $\sim$ 10\% and $\sim$ 3\%, respectively. These values use rough estimates of the host stellar mass, based on integrated magnitudes and a fixed mass-to-light ratio. These ratios are within the scatter for galaxies of similar mass  \citep{Ordenes2018, SanchezJanssen2019}. While the NSC-to-host mass relation appears to be independent of environment \citep{SanchezJanssen2019}, the mass ratio itself shows a trend with host stellar mass. The mass ratio seems to be higher in lower mass galaxies, similar to what we find in the NSC-host system in the lower mass galaxy KKs58 compared to the more massive galaxy KK197. Nucleated late-type dwarf galaxies in low density environments can show a similar NSC-to-host mass fraction \citep{Georgiev2016}.

In our sample of 14 dwarf galaxies studied with MUSE, we found two nucleated dwarf galaxies. With a more complete sample, the effects of the Centaurus group environment on the nucleation fraction could be studied.
\cite{SanchezJanssen2019} presented a study of the nucleation fraction of galaxies in different environments. They suggest that the nucleation fraction of dwarf galaxies appears to be lower in low density environments such as the Local Group when compared to the higher density environments of the Fornax, Virgo, or Coma clusters. The environment might also have an affect on the limiting galaxy mass threshold below which the nucleation fraction drops to zero. Studying dwarfs in the Fornax cluster, \cite{Ordenes2018} determined a limiting mass of $\sim 2.5 \times 10^6 M_\sun$, while \cite{SanchezJanssen2019} identified a limiting mass of $\sim 5 \times 10^5 M_\sun$ in their study of the core region of the Virgo cluster. With a stellar mass of $7.3 \times 10^6 M_\sun$, KKs58 is already close to these thresholds. It is unclear if this limiting mass arises because low mass galaxies become less efficient in forming NSCs or if there is a mechanism that can destroy them more easily. Due to its proximity, the Centaurus group provides an ideal laboratory to study the environmental effects on the nucleation fraction and the dwarf galaxies themselves. 

%--------------------- NSC formation
\subsection{Insights on nuclear star cluster formation}
\label{sect:dis_NSC}
The formation of NSCs is still a heavily debated issue. As described in Sect. \ref{sect:intro}, typically two pathways are discussed: the in-situ formation directly at the galaxy's centre and formation from the in-spiral and merging of GCs. With the collected data of KKs58 and KK197, we can put constraints on the formation of their NSCs. 

The NSCs of both KKs58 and KK197 have a relatively simple star formation history with only one peak at $\sim 7$ Gyr. We could not find any younger populations or emission lines that would indicate ongoing star formation. Nonetheless, our rough estimates of the ages of the host galaxies from their low S/N spectra ($<$ 50 \AA$^{-1}$) indicate that the NSC in KK197 could be younger than the main stellar body. However, measuring accurate ages of old ($>$ 5 Gyr) stellar populations with integrated spectra is generally challenging (e.g. \citealt{Spengler2017, Usher2019}), and even more so in this case because the MUSE instrument lacks important age-sensitive features such as higher-order Balmer lines at bluer wavelengths. In addition, while the host galaxy spectra have only a low S/N, even the age measurements from the high S/N NSC spectra might be biased towards younger ages due to the presence of hot horizontal branch (HB) stars that most likely are present in massive star clusters \citep{Schiavon2004, Ocvirk2010, Georgiev2012}. In the MW, almost all massive GCs have hot HB stars \citep{RecioBlanco2006, Lee2007} and their high effective temperatures ($T_\text{eff} \sim 9000$ K) lead to a bias in spectroscopic metallicities towards younger ages especially in blue, metal-poor GCs of $\sim$ 4 Gyr compared to isochrone fitting (e.g. for NGC\,3201 and NGC\,5024 \citealt{Dotter2010, Perina2011}). Consequently, the age difference between NSC and host found in KK197 might be the result from uncertainty in the full spectral fitting and an additional bias introduced by hot HB stars. Due to stellar crowding in the NSC, we cannot use isochrone fitting to the HST data to obtain a photometric age. 

We observed a clear difference in metallicity between the two NSCs and their host galaxies. In KK197, we found a mean host galaxy metallicity of [Fe/H] = $-0.84$ dex, while the NSC and the bright GC (KK197-03) have a metallicity of [Fe/H] = $-1.84$ dex. The NSC in KKs58 has a similarly low metallicity of [Fe/H] $\sim$ $-1.80$ dex and again, the host galaxy appears to be more metal-rich with a mean metallicity of $\sim -1.35$ dex. For both galaxies, our spectroscopic metallicity estimates are in agreement with photometric estimates \citep{Crnojevic2010, Mueller2019} and both galaxies follow the mass-metallicity relation of dwarf galaxies \citep{Kirby2013}.

That the NSCS are more metal-poor than their hosts is in contrast to the study by \cite{Paudel2011}, who investigated nuclei of dwarf ellipticals in the Virgo cluster and established that a majority of them are more metal-rich then their host. However, these nuclei have absolute $r$-band magnitudes $< -11$ mag and are thus significantly more luminous than the NSCs of KKs58 and KK197. Since these NCSs are brighter and likely more massive than those studied here, a higher metallicity is fully expected \citep{Spengler2017}. 

The simple star formation history and the NSCs being more metal-poor than their hosts disagrees with the in-situ formation of the NSCs in KKs58 and KK197. The in-situ scenario typically favours a more metal-rich NSC due to fast, repetitive gas accretion from the host's gas reservoir (e.g. \citealt{Milosavljevic2004, Bekki2007}) and a prolonged star formation history at the centre of the potential well \citep{Antonini2015}.
Because we find that the two NSCs have significantly lower metallicities than their hosts globally, it is unlikely that they have formed at the centres alongside with the metal-rich host galaxy, otherwise they should have experienced a complex star formation history and would be polluted by metal-richer stars.
When considering inflow of very metal-poor gas to the centre, for example from cold filaments \citep{Cresci2010}, we would expect to see an additional metal-rich population and an extended star formation history.

The GC accretion scenario, therefore, is more likely in the case of KKs58 and KK197. In this scenario, the NSC forms by accreting gas-free GCs that spiral into the galaxy's centre due to dynamical friction \citep[e.g.][]{Tremaine1975, ArcaSedda2014}. Consequently, the formed NSC should reflect the properties of the accreted GCs that are typically more metal-poor than their hosts (e.g. \citealt{Lamers2017}). In KK197 this is further supported by the finding of the bright GC close in projection to the centre that shares the low metallicity of the NSC. In case the velocity offset between NSC and host in KK197 is confirmed, the NSC might not have settled at the centre yet and could still be in the process of spiralling in. 

The presence of at least two other GCs in KK197 that have not sunk to the centre could be explained by larger initial formation distances \citep{Angus2009}, but might also provide insight into the underlying dark matter (DM) distribution of the host galaxy. As has been discussed extensively for the case of the GC system of the Fornax dwarf spheroidal galaxy (e.g. \citealt{Goerdt2006, Boldrini2019, Meadows2019}), the orbits of in-spiralling GCs are affected by the underlying DM distribution and could be used to differentiate between cuspy or cored DM profiles. GCs in dwarf galaxies are expected to quickly spiral towards the centre in cuspy DM halos (e.g. \citealt{Tremaine1975}), but might stall near the core radius in cored DM halos \citep{Goerdt2006}. Constraining the inner DM slope observationally is challenging because of degeneracies with orbital anisotropy, mass-to-light ratios, halo shapes and initial formation location of the GCs. With our limited data of KK197, we cannot study the underlying DM profile, but with precise ages and masses of the star clusters and the host galaxy, KK197 would provide an interesting testbed to study the cusp-core problem.
The possible age difference between NSC and host we observed in KK197, if confirmed, could also resolve this so called timing problem, if the star clusters formed later at larger distances from metal-poor gas. A similar scenario has been recently discussed for the Fornax dwarf spheroidal galaxy \citep{Leung2019}.

We could not identify GCs in KKs58. All possible candidates turned out to be either foreground stars of the MW or background galaxies. Although we cannot exclude that there are undetected GCs outside the MUSE FOV, it is also possible that KKs58 only has one star cluster, the NSC. Under the assumption that the NSC was purely formed out of GCs, the NSC would then contain the entire original GC population of KKs58 with the exception of some that might have been stripped in tidal interactions with the group environment or are outside of the MUSE FOV.
To test whether this is reasonable, we can compare the NSC-to-host mass ratio of $\sim 10$\% to the GC system-to-host mass ratio typically found in such low mass galaxies. \mbox{\cite{Forbes2018}} studied the GC system-to-host mass relation over a large range of host masses and found that the ratio can scatter between 0.1 and 10\%. Consequently, KKs58 is already within this scatter without the presence of additional star clusters.
The same conclusion can be drawn when comparing the GC specific frequency $S_N$\footnote{$S_N = N_{\text{GC}} \times 10^{0.4(M_V + 15)}$, \citep{Harris1981}} of KKs58 to other dwarf galaxies. Assuming the NSC is of GC origin, KKs58 has $S_N \sim 16$ and KK197, with three confirmed star clusters, has $S_N \sim 18$. Both are in agreement with other dwarf galaxies of similar brightness that have $S_N \approx$ 10 -- 20 \citep{Georgiev2010}.

\subsection{Possible UCD progenitors?}
\label{sect:UCD_dis}
\begin{figure}
\centering
\includegraphics[width=0.49\textwidth]{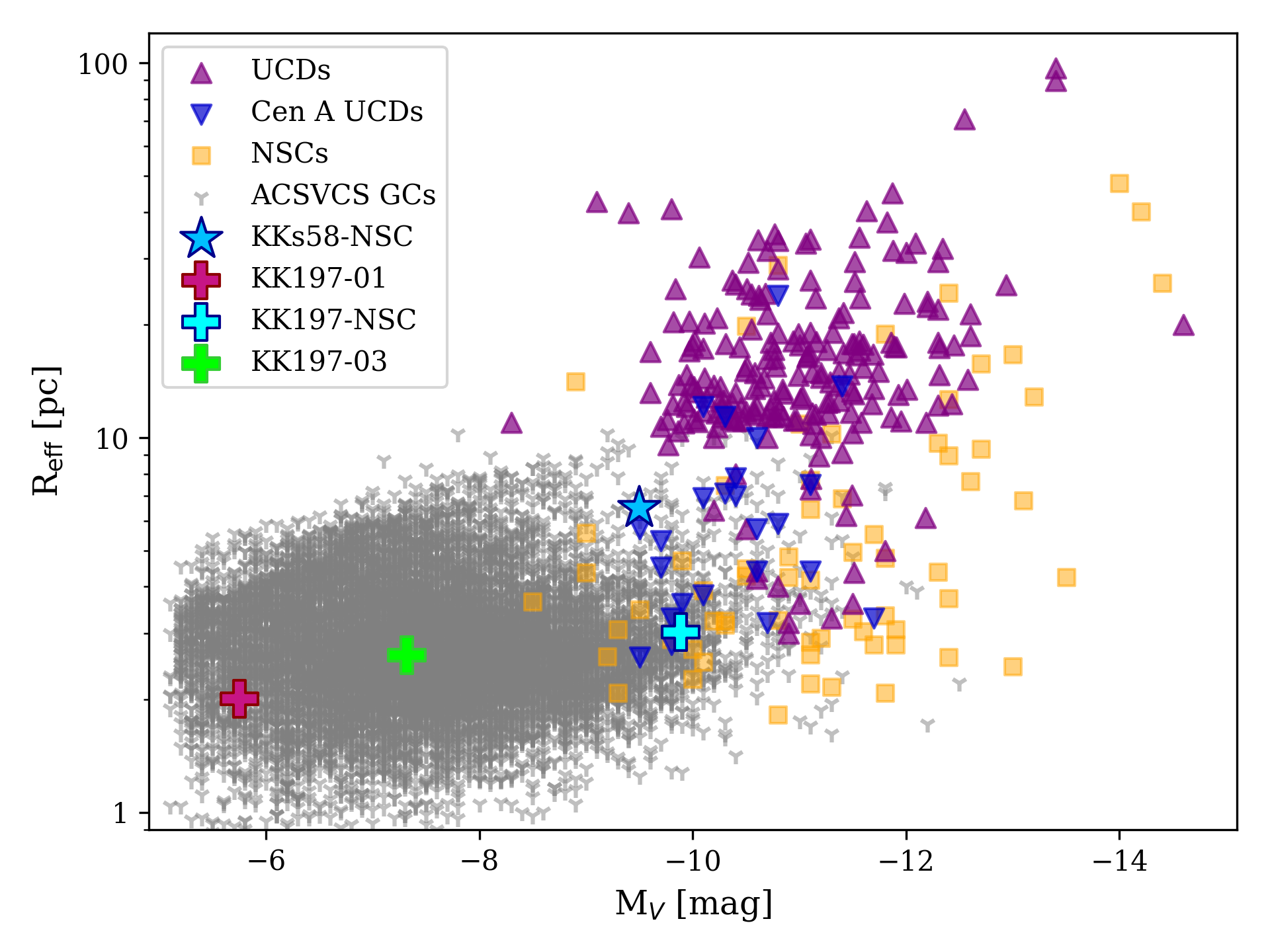}
\caption{$V$-band magnitudes versus effective radius for different compact stellar systems. NSCs and star clusters compilations are taken from \cite{Misgeld2011}, the UCDs from \cite{Fahrion2019}. We highlight the UCDs of Cen A with blue triangles. The NSC of KKs58 is shown with the blue star. The star clusters of KK197 are shown by the crosses.}
\label{fig:mag_vs_rad}
\end{figure}

We compare the NSC of KKs58 and the star clusters of KK197 to other compact stellar systems in Fig. \ref{fig:mag_vs_rad}. This figure shows effective radii and absolute $V$-band magnitudes for a large sample of GCs, UCDs, and NSCs from literature compilations presented in \cite{Misgeld2011} and \cite{Fahrion2019}. We highlight the UCDs associated with Cen A \citep{Rejkuba2007} to illustrate that those are generally on the fainter side of the UCD population and some even have a very similar size and magnitude as the NSCs of KKs58 and KK197. The NSCs are also placed among the fainter NSC population, especially the NSC of KK197. The two GCs in KK197 are well placed among other faint GCs known from studies of galaxy cluster environments.

In case KKs58 or KK197 would get destroyed in the group environment and the NSCs would be stripped without being destroyed or significantly altered, the remnant NSCs would fit within the UCD population, but might also be interpreted as a GC due to their magnitude and low mass. This highlights the ambiguity connected to the question of origin for UCDs. It has been suggested that UCDs could be the stripped nuclei of disrupted galaxies \citep{Phillipps2001, Bekki2003, Drinkwater2003, Pfeffer2013, Strader2013}, the high-mass end of the star cluster population \citep{Mieske2002, Mieske2004, Kissler-Patig2006}, or the result of merged star clusters \citep{Fellhauer2002, Maraston2004, Fellhauer2005}. So far, the only unambiguous confirmations of stripped NSC-type UCDs have been made for the most massive, metal-rich UCDs using either the presence of a SMBH detected with high-resolution IFU observations \citep{Seth2014, Ahn2017, Ahn2018, Afanasiev2018} or from the detection of a complex, extended star formation history \citep{Norris2015}. In contrast, objects like M\,54 and $\omega$\,Cen in the MW, as well as the comparison of the NSCs of KKs58 and KK197 to Cen A UCDs, show that the population of low-mass UCDs most likely also contains a significant population of stripped NSCs. 
Using simulations, \cite{Pfeffer2014} predicted the number of UCDs from stripped NSCs of a given stellar mass as a function of environment virial mass. For the Centaurus group, virial masses between 4.0 $\times 10^{12}$ and 1.4 $\times 10^{13} M_\sun$ are discussed \citep{Karachentsev2007, Woodley2006, Woodley2007, Woodley2010} and due to this large scatter, the number of predicted UCDs from stripped NSCs with masses $> 10^5 M_\sun$ ranges between $\sim$ 4 -- 13, following the prescription of \cite{Pfeffer2014}. Because these predictions strongly depend on the nucleation fraction of galaxies, assessing the number and masses of dwarf galaxy NSCs can help to improve such models. 

Confirming the NSC origin of a low-mass, metal-poor UCD is particularly challenging because its properties are mostly indistinguishable from a high-mass GC (see also \citealt{Fahrion2019}). In the case of KKs58, the extended size and the ellipticity could give an indication, if it were to survive unaltered the disruption of its host. On the other hand, in case KK197 would be stripped of its NSC, it would be very difficult to distinguish it from the general intra-group population of low-mass GCs. However, simulations by \cite{Pfeffer2013} have shown that while the original NSCs do not expand during the stripping process, they can retain an remnant envelope from the galaxy body causing the resulting UCD to appear extended. Therefore, extended sizes of GCs at these low masses and metallicities could give further evidence of NSC-origin. Examples of faint envelopes around compact GC-like objects have been found in the Virgo \citep{Liu2015} as well as in the Fornax cluster \citep{Voggel2016}.

The elevated dynamical mass compared to our stellar population estimate of KK197-NSC is another observable that this NSC shares with many UCDs and some GCs \citep{Hasegan2005, Rejkuba2007, Mieske2008, Taylor2010}. \cite{Mieske2013} found a mean fraction of dynamical-to-stellar mass-to-light ratios of 1.7 for massive UCDs, which is lower than our result for KK197-NSC, but still consistent within the scatter. As discussed before, our stellar population properties of the NSCs might be biased to lower ages due to the presence of hot horizontal branch stars. This also affects the $M/L$ we use for the mass determination. For example, assuming an age of 13 Gyr instead of the best-fit age of 7 Gyr at the metallicity of KK197-NSC, increases the $M/L_V$ from 1.4 to 2.1. In addition, our mass modelling assumes spherical symmetry and isotropy, and does not account for internal rotation of the NSC. However, simulations and observations have shown that NSCs can have complex density profiles \citep{Boeker2002b} and kinematics (e.g. \citealt{Lyubenova2013, Perets2014, Lyubenova2019, Fahrion2019b}).

The elevated dynamical $M/L$ can also have physical origin. Variations of the IMF in the clusters can result in this difference, both when considering top-heavy \citep{Murray2009, Dabringhausen2009} and bottom-heavy IMFs \citep{MieskeKroupa2008}, and a central SMBH can also increase the dynamical $M/L$ \citep{Mieske2013}.

%------------------------- Conclusions ---------------------
\section{Conclusions}
\label{sect:conclusions}
In this paper, we report the discovery of a NSC in KKs58, a dwarf galaxy member of the Centaurus group, and analyse its properties based on new MUSE data and ancillary photometric data from FORS2 and DECam. Furthermore, we analyse the NSC and two GCs in KK197, another dwarf galaxy in the Centaurus group that was previously studied photometrically with HST ACS data \citep{Georgiev2009b}. We summarise our results as follows:
\begin{itemize}
\item{We fitted KKs58's DECam $g$-band surface brightness image using a double S\'{e}rsic profile. We determined the NSC to be elliptical with an effective radius of $6.6\pm 0.5$ pc at an absolute magnitude of $M_g$ = $-$9.19 mag ($M_V = -9.51 \pm 0.07$ mag).}
\item{We extracted a high S/N MUSE spectrum of KKs58-NSC and measured its LOS velocity (v = 474.6 $\pm$ 1.9 km s$^{-1}$) and stellar population properties with full spectral fitting. The NSC is at least $\sim$ 7 Gyr old and metal poor ([Fe/H] = $-1.75$ dex). Using the stellar population analysis, we obtain a stellar mass of $M_{\ast_\mathrm{,KKs58-NSC}} = 7.3 \times 10^5 M_\sun$.}
\item{The low surface brightness of KKs58 itself does not allow to obtain a map of its kinematic or stellar population properties. Instead, we extracted a single, low S/N spectrum and measured the LOS velocity of v = 482.6 $\pm$ 12.6 km s$^{-1}$. The low S/N gives only rough estimates of the age of $\sim$ 7 Gyr and metallicity of [Fe/H] $\sim$ $-1.35$ dex. The photometric metallicity estimate \citep{Mueller2019} is consistent with the NSC metallicity. For the host galaxy, we estimated the stellar mass of $M_{\ast_\mathrm{, KKs58}} \approx 7 \times 10^6 M_\sun$.}
\item{We confirmed the membership of three star clusters in KK197 based on their radial velocities. The stellar density is high enough to obtain a Voronoi binned map of the radial velocities and metallicity distribution. The galaxy shows a rotation signature with a projected maximum amplitude of $\sim$ 5 km s$^{-1}$.}
\item{The NSC of KK197 might show an offset velocity to the host galaxy of \mbox{$\Delta v = 7.8 \pm 3.8$ km s$^{-1}$}. Furthermore, its metallicity of [Fe/H] = $-1.84$ dex is significantly lower than the surrounding galaxy field stellar population ([Fe/H] $\sim$ $-$0.84 dex). Using a high-resolution VLT UVES spectrum of KK197-NSC, we determined the dynamical mass of this star cluster to be $M_{\text{dyn, KK197-NSC}} \approx 4 \times 10^6 M_\sun$, while we find a stellar mass of \mbox{$\sim 1 \times 10^6 M_\sun$.}}
\item{We found for both dwarf galaxies that the NSCs are more metal-poor than the host galaxies. Comparing this to basic expectations from different NSC formation scenarios, the lower metallicity in the NSC makes a strong case for the GC accretion scenario, where the NSC forms out of inspiraling GCs that were formed farther out in the galaxy. The metal-poor GC found in KK197 further supports this scenario.}
\item{We estimated NSC mass-to-host mass ratios of $\sim$ $3 - 10$\% in the two dwarf galaxies, typical values for low-mass dwarf galaxies. The masses of the NSCs are also typical for NSCs of a dwarf galaxy as the comparison to larger samples of nucleated dwarfs shows. However, the stellar mass of KKs58 places it in a mass range, where typically a very low nucleation fraction is observed.}
\item{The mass, size and metallicity places the NSCs of KKs58 and KK197 within the range of 
other NSCs, but also among known low-mass UCDs of the Centaurus group. This suggests
that some of these UCDs might have originated from stripped NSCs of disrupted dwarf galaxies like KKs58 or KK197, although their observed properties today make them indistinguishable from the high mass GC population.}
\end{itemize}

Out of a sample of 14 dwarf galaxies observed with MUSE, we were able to identify two nucleated dwarfs. Although there have been successful attempts in searching for NSCs in dwarf galaxies of the Centaurus group (e.g. \citealt{Georgiev2009a}), the nucleation fraction of this environment is still unknown. Identifying potential NSCs in low-mass dwarf galaxies is generally challenging and requires either a spectroscopic measurement to confirm the membership to a galaxy via radial velocities or high-resolution photometry. At the same time, extracting even a single spectrum from an MUSE observation of faint host galaxies to measure radial velocities or stellar population properties is also difficult. In these cases, GCs or even NSCs provide excellent tracers of otherwise inaccessible properties of the host galaxies.

\begin{acknowledgements}
We thank the anonymous referee for helpful comments that helped to improve this manuscript.
This work is based on observations collected at the European Organization for Astronomical Research in the Southern Hemisphere under ESO programmes 0101.A-0193, 084.D-0818, 069.D-0169. KF is grateful to Nadine Neumayer and Alina B\"ocker for helpful discussions. OM is grateful to the Swiss National Science Foundation for financial support. GvdV acknowledges funding from the European Research Council (ERC) under the European Union's Horizon 2020 research and innovation programme under grant agreement No 724857 (Consolidator Grant ArcheoDyn). HJ acknowledges support from the Australian Research Council through the Discovery Project DP150100862. This research made use of Photutils, an Astropy package for detection and photometry of astronomical sources \citep{Bradley2019}. This research made use of Astropy,\footnote{http://www.astropy.org} a community-developed core Python package for Astronomy \citep{astropy:2013, astropy:2018}. This research has made use of the NASA/IPAC Extragalactic Database (NED), which is funded by the National Aeronautics and Space Administration and operated by the California Institute of Technology. 
\end{acknowledgements}

\bibliographystyle{aa} % style aa.bst
\bibliography{References}

\begin{thebibliography}{163}
\expandafter\ifx\csname natexlab\endcsname\relax\def\natexlab#1{#1}\fi

\bibitem[{{Afanasiev} {et~al.}(2018){Afanasiev}, {Chilingarian}, {Mieske},
  {Voggel}, {Picotti}, {Hilker}, {Seth}, {Neumayer}, {Frank}, {Romanowsky},
  {Hau}, {Baumgardt}, {Ahn}, {Strader}, {den Brok}, {McDermid}, {Spitler},
  {Brodie}, \& {Walsh}}]{Afanasiev2018}
{Afanasiev}, A.~V., {Chilingarian}, I.~V., {Mieske}, S., {et~al.} 2018, \mnras,
  477, 4856

\bibitem[{{Agarwal} \& {Milosavljevi{\'c}}(2011)}]{Agarwal2011}
{Agarwal}, M. \& {Milosavljevi{\'c}}, M. 2011, \apj, 729, 35

\bibitem[{{Ahn} {et~al.}(2018){Ahn}, {Seth}, {Cappellari}, {Krajnovi{\'c}},
  {Strader}, {Voggel}, {Walsh}, {Bahramian}, {Baumgardt}, {Brodie},
  {Chilingarian}, {Chomiuk}, {den Brok}, {Frank}, {Hilker}, {McDermid},
  {Mieske}, {Neumayer}, {Nguyen}, {Pechetti}, {Romanowsky}, \&
  {Spitler}}]{Ahn2018}
{Ahn}, C.~P., {Seth}, A.~C., {Cappellari}, M., {et~al.} 2018, \apj, 858, 102

\bibitem[{{Ahn} {et~al.}(2017){Ahn}, {Seth}, {den Brok}, {Strader},
  {Baumgardt}, {van den Bosch}, {Chilingarian}, {Frank}, {Hilker}, {McDermid},
  {Mieske}, {Romanowsky}, {Spitler}, {Brodie}, {Neumayer}, \&
  {Walsh}}]{Ahn2017}
{Ahn}, C.~P., {Seth}, A.~C., {den Brok}, M., {et~al.} 2017, \apj, 839, 72

\bibitem[{{Anderson} \& {King}(2000)}]{Anderson2000}
{Anderson}, J. \& {King}, I.~R. 2000, \pasp, 112, 1360

\bibitem[{{Angus} \& {Diaferio}(2009)}]{Angus2009}
{Angus}, G.~W. \& {Diaferio}, A. 2009, \mnras, 396, 887

\bibitem[{{Antonini}(2013)}]{Antonini2013}
{Antonini}, F. 2013, \apj, 763, 62

\bibitem[{Antonini(2014)}]{Antonini2014}
Antonini, F. 2014, The Astrophysical Journal, 794, 106

\bibitem[{{Antonini} {et~al.}(2015){Antonini}, {Barausse}, \&
  {Silk}}]{Antonini2015}
{Antonini}, F., {Barausse}, E., \& {Silk}, J. 2015, \apj, 812, 72

\bibitem[{{Arca-Sedda} \& {Capuzzo-Dolcetta}(2014)}]{ArcaSedda2014}
{Arca-Sedda}, M. \& {Capuzzo-Dolcetta}, R. 2014, \apj, 785, 51

\bibitem[{{Astropy Collaboration} {et~al.}(2013){Astropy Collaboration},
  {Robitaille}, {Tollerud}, {Greenfield}, {Droettboom}, {Bray}, {Aldcroft},
  {Davis}, {Ginsburg}, {Price-Whelan}, {Kerzendorf}, {Conley}, {Crighton},
  {Barbary}, {Muna}, {Ferguson}, {Grollier}, {Parikh}, {Nair}, {Unther},
  {Deil}, {Woillez}, {Conseil}, {Kramer}, {Turner}, {Singer}, {Fox}, {Weaver},
  {Zabalza}, {Edwards}, {Azalee Bostroem}, {Burke}, {Casey}, {Crawford},
  {Dencheva}, {Ely}, {Jenness}, {Labrie}, {Lim}, {Pierfederici}, {Pontzen},
  {Ptak}, {Refsdal}, {Servillat}, \& {Streicher}}]{astropy:2013}
{Astropy Collaboration}, {Robitaille}, T.~P., {Tollerud}, E.~J., {et~al.} 2013,
  \aap, 558, A33

\bibitem[{{Bacon} {et~al.}(2010){Bacon}, {Accardo}, {Adjali}, {Anwand},
  {Bauer}, {Biswas}, {Blaizot}, {Boudon}, {Brau-Nogue}, {Brinchmann},
  {Caillier}, {Capoani}, {Carollo}, {Contini}, {Couderc}, {Daguis{\'e}},
  {Deiries}, {Delabre}, {Dreizler}, {Dubois}, {Dupieux}, {Dupuy}, {Emsellem},
  {Fechner}, {Fleischmann}, {Fran{\c c}ois}, {Gallou}, {Gharsa}, {Glindemann},
  {Gojak}, {Guiderdoni}, {Hansali}, {Hahn}, {Jarno}, {Kelz}, {Koehler},
  {Kosmalski}, {Laurent}, {Le Floch}, {Lilly}, {Lizon}, {Loupias}, {Manescau},
  {Monstein}, {Nicklas}, {Olaya}, {Pares}, {Pasquini}, {P{\'e}contal-Rousset},
  {Pell{\'o}}, {Petit}, {Popow}, {Reiss}, {Remillieux}, {Renault}, {Roth},
  {Rupprecht}, {Serre}, {Schaye}, {Soucail}, {Steinmetz}, {Streicher}, {Stuik},
  {Valentin}, {Vernet}, {Weilbacher}, {Wisotzki}, \& {Yerle}}]{Bacon2010}
{Bacon}, R., {Accardo}, M., {Adjali}, L., {et~al.} 2010, in \procspie, Vol.
  7735, Ground-based and Airborne Instrumentation for Astronomy III, 773508

\bibitem[{{Baumgardt} \& {Hilker}(2018)}]{BaumgardtHilker2018}
{Baumgardt}, H. \& {Hilker}, M. 2018, \mnras, 478, 1520

\bibitem[{{Bekki}(2007)}]{Bekki2007}
{Bekki}, K. 2007, Publications of the Astronomical Society of Australia, 24, 77

\bibitem[{{Bekki} {et~al.}(2003){Bekki}, {Couch}, {Drinkwater}, \&
  {Shioya}}]{Bekki2003}
{Bekki}, K., {Couch}, W.~J., {Drinkwater}, M.~J., \& {Shioya}, Y. 2003, \mnras,
  344, 399

\bibitem[{{Bekki} {et~al.}(2006){Bekki}, {Couch}, \& {Shioya}}]{Bekki2006}
{Bekki}, K., {Couch}, W.~J., \& {Shioya}, Y. 2006, \apjl, 642, L133

\bibitem[{{Bellazzini} {et~al.}(2008){Bellazzini}, {Ibata}, {Chapman},
  {Mackey}, {Monaco}, {Irwin}, {Martin}, {Lewis}, \&
  {Dalessandro}}]{Bellazzini2008}
{Bellazzini}, M., {Ibata}, R.~A., {Chapman}, S.~C., {et~al.} 2008, \aj, 136,
  1147

\bibitem[{{Boecker} {et~al.}(2019){Boecker}, {Leaman}, {van de Ven}, {Norris},
  {Mackereth}, \& {Crain}}]{Boecker2019}
{Boecker}, A., {Leaman}, R., {van de Ven}, G., {et~al.} 2019, \mnras, 2678

\bibitem[{{B{\"o}ker} {et~al.}(2004{\natexlab{a}}){B{\"o}ker}, {Sarzi},
  {McLaughlin}, {van der Marel}, {Rix}, {Ho}, \& {Shields}}]{Boker2004}
{B{\"o}ker}, T., {Sarzi}, M., {McLaughlin}, D.~E., {et~al.} 2004{\natexlab{a}},
  \aj, 127, 105

\bibitem[{{B{\"o}ker} {et~al.}(2004{\natexlab{b}}){B{\"o}ker}, {Sarzi},
  {McLaughlin}, {van der Marel}, {Rix}, {Ho}, \& {Shields}}]{Boeker2002b}
{B{\"o}ker}, T., {Sarzi}, M., {McLaughlin}, D.~E., {et~al.} 2004{\natexlab{b}},
  \aj, 127, 105

\bibitem[{{Boldrini} {et~al.}(2019){Boldrini}, {Mohayaee}, \&
  {Silk}}]{Boldrini2019}
{Boldrini}, P., {Mohayaee}, R., \& {Silk}, J. 2019, \mnras, 485, 2546

\bibitem[{Bradley {et~al.}(2019)Bradley, Sipocz, Robitaille, Tollerud,
  Vinícius, Deil, Barbary, Busko, Günther, Cara, Wilson, Conseil, Droettboom,
  Bostroem, Bray, Bratholm, Lim, Craig, Barentsen, Pascual, Donath, Greco,
  Perren, Kerzendorf, de~Val-Borro, Dencheva, de~Albernaz~Ferreira, Souchereau,
  D'Eugenio, \& Weaver}]{Bradley2019}
Bradley, L., Sipocz, B., Robitaille, T., {et~al.} 2019, astropy/photutils:
  v0.7.1

\bibitem[{{Cappellari}(2017)}]{Cappellari2017}
{Cappellari}, M. 2017, \mnras, 466, 798

\bibitem[{{Cappellari} \& {Copin}(2003)}]{CappellariCopin2003}
{Cappellari}, M. \& {Copin}, Y. 2003, \mnras, 342, 345

\bibitem[{{Cappellari} \& {Emsellem}(2004)}]{Cappellari2004}
{Cappellari}, M. \& {Emsellem}, E. 2004, \pasp, 116, 138

\bibitem[{{Capuzzo-Dolcetta}(1993)}]{CapuzzoDolcetta1993}
{Capuzzo-Dolcetta}, R. 1993, \apj, 415, 616

\bibitem[{{Capuzzo-Dolcetta} \& {Miocchi}(2008)}]{CapuzzoDolcetta2008}
{Capuzzo-Dolcetta}, R. \& {Miocchi}, P. 2008, \apj, 681, 1136

\bibitem[{{C{\^o}t{\'e}} {et~al.}(2006){C{\^o}t{\'e}}, {Piatek}, {Ferrarese},
  {Jord{\'a}n}, {Merritt}, {Peng}, {Ha{\c s}egan}, {Blakeslee}, {Mei}, {West},
  {Milosavljevi{\'c}}, \& {Tonry}}]{Cote2006}
{C{\^o}t{\'e}}, P., {Piatek}, S., {Ferrarese}, L., {et~al.} 2006, \apjs, 165,
  57

\bibitem[{{Cresci} {et~al.}(2010){Cresci}, {Mannucci}, {Maiolino}, {Marconi},
  {Gnerucci}, \& {Magrini}}]{Cresci2010}
{Cresci}, G., {Mannucci}, F., {Maiolino}, R., {et~al.} 2010, \nat, 467, 811

\bibitem[{{Crnojevi{\'c}} {et~al.}(2010){Crnojevi{\'c}}, {Grebel}, \&
  {Koch}}]{Crnojevic2010}
{Crnojevi{\'c}}, D., {Grebel}, E.~K., \& {Koch}, A. 2010, \aap, 516, A85

\bibitem[{{Crnojevi{\'c}} {et~al.}(2019){Crnojevi{\'c}}, {Sand}, {Bennet},
  {Pasetto}, {Spekkens}, {Caldwell}, {Guhathakurta}, {McLeod}, {Seth}, {Simon},
  {Strader}, \& {Toloba}}]{Crnojevic2019}
{Crnojevi{\'c}}, D., {Sand}, D.~J., {Bennet}, P., {et~al.} 2019, \apj, 872, 80

\bibitem[{{Crnojevi{\'c}} {et~al.}(2014){Crnojevi{\'c}}, {Sand}, {Caldwell},
  {Guhathakurta}, {McLeod}, {Seth}, {Simon}, {Strader}, \&
  {Toloba}}]{Crnojevic2014}
{Crnojevi{\'c}}, D., {Sand}, D.~J., {Caldwell}, N., {et~al.} 2014, \apjl, 795,
  L35

\bibitem[{{Crnojevi{\'c}} {et~al.}(2016){Crnojevi{\'c}}, {Sand}, {Spekkens},
  {Caldwell}, {Guhathakurta}, {McLeod}, {Seth}, {Simon}, {Strader}, \&
  {Toloba}}]{Crnojevic2016}
{Crnojevi{\'c}}, D., {Sand}, D.~J., {Spekkens}, K., {et~al.} 2016, \apj, 823,
  19

\bibitem[{{Da Rocha} {et~al.}(2011){Da Rocha}, {Mieske}, {Georgiev}, {Hilker},
  {Ziegler}, \& {Mendes de Oliveira}}]{DaRocha2011}
{Da Rocha}, C., {Mieske}, S., {Georgiev}, I.~Y., {et~al.} 2011, \aap, 525, A86

\bibitem[{{Dabringhausen} {et~al.}(2009){Dabringhausen}, {Kroupa}, \&
  {Baumgardt}}]{Dabringhausen2009}
{Dabringhausen}, J., {Kroupa}, P., \& {Baumgardt}, H. 2009, \mnras, 394, 1529

\bibitem[{{Dekker} {et~al.}(2000){Dekker}, {D'Odorico}, {Kaufer}, {Delabre}, \&
  {Kotzlowski}}]{Dekker2000}
{Dekker}, H., {D'Odorico}, S., {Kaufer}, A., {Delabre}, B., \& {Kotzlowski}, H.
  2000, Society of Photo-Optical Instrumentation Engineers (SPIE) Conference
  Series, Vol. 4008, {Design, construction, and performance of UVES, the
  echelle spectrograph for the UT2 Kueyen Telescope at the ESO Paranal
  Observatory}, ed. M.~{Iye} \& A.~F. {Moorwood}, 534--545

\bibitem[{{den Brok} {et~al.}(2014){den Brok}, {Peletier}, {Seth}, {Balcells},
  {Dominguez}, {Graham}, {Carter}, {Erwin}, {Ferguson}, {Goudfrooij},
  {Guzm{\'a}n}, {Hoyos}, {Jogee}, {Lucey}, {Phillipps}, {Puzia}, {Valentijn},
  {Verdoes Kleijn}, \& {Weinzirl}}]{denBrok2014}
{den Brok}, M., {Peletier}, R.~F., {Seth}, A., {et~al.} 2014, \mnras, 445, 2385

\bibitem[{{Dotter} {et~al.}(2010){Dotter}, {Sarajedini}, {Anderson},
  {Aparicio}, {Bedin}, {Chaboyer}, {Majewski}, {Mar{\'\i}n-Franch}, {Milone},
  {Paust}, {Piotto}, {Reid}, {Rosenberg}, \& {Siegel}}]{Dotter2010}
{Dotter}, A., {Sarajedini}, A., {Anderson}, J., {et~al.} 2010, \apj, 708, 698

\bibitem[{{Drinkwater} {et~al.}(2003){Drinkwater}, {Gregg}, {Hilker}, {Bekki},
  {Couch}, {Ferguson}, {Jones}, \& {Phillipps}}]{Drinkwater2003}
{Drinkwater}, M.~J., {Gregg}, M.~D., {Hilker}, M., {et~al.} 2003, \nat, 423,
  519

\bibitem[{{Eigenthaler} {et~al.}(2018){Eigenthaler}, {Puzia}, {Taylor},
  {Ordenes-Brice{\~n}o}, {Mu{\~n}oz}, {Ribbeck}, {Alamo-Mart{\'\i}nez},
  {Zhang}, {{\'A}ngel}, {Capaccioli}, {C{\^o}t{\'e}}, {Ferrarese}, {Galaz},
  {Grebel}, {Hempel}, {Hilker}, {Lan{\c{c}}on}, {Mieske}, {Miller}, {Paolillo},
  {Powalka}, {Richtler}, {Roediger}, {Rong}, {S{\'a}nchez-Janssen}, \&
  {Spengler}}]{Eigenthaler2018}
{Eigenthaler}, P., {Puzia}, T.~H., {Taylor}, M.~A., {et~al.} 2018, \apj, 855,
  142

\bibitem[{{Erwin}(2015)}]{Erwin2015}
{Erwin}, P. 2015, \apj, 799, 226

\bibitem[{{Fahrion} {et~al.}(2019{\natexlab{a}}){Fahrion}, {Georgiev},
  {Hilker}, {Lyubenova}, {van de Ven}, {Alfaro-Cuello}, {Corsini}, {Sarzi},
  {McDermid}, \& {de Zeeuw}}]{Fahrion2019}
{Fahrion}, K., {Georgiev}, I., {Hilker}, M., {et~al.} 2019{\natexlab{a}}, \aap,
  625, A50

\bibitem[{{Fahrion} {et~al.}(2019{\natexlab{b}}){Fahrion}, {Lyubenova}, {van de
  Ven}, {Leaman}, {Hilker}, {Mart{\'\i}n-Navarro}, {Zhu}, {Alfaro-Cuello},
  {Coccato}, {Corsini}, {Falc{\'o}n-Barroso}, {Iodice}, {McDermid}, {Sarzi}, \&
  {de Zeeuw}}]{Fahrion2019b}
{Fahrion}, K., {Lyubenova}, M., {van de Ven}, G., {et~al.} 2019{\natexlab{b}},
  \aap, 628, A92

\bibitem[{{Falc{\'o}n-Barroso} {et~al.}(2011){Falc{\'o}n-Barroso},
  {S{\'a}nchez-Bl{\'a}zquez}, {Vazdekis}, {Ricciardelli}, {Cardiel}, {Cenarro},
  {Gorgas}, \& {Peletier}}]{FalconBarroso2011}
{Falc{\'o}n-Barroso}, J., {S{\'a}nchez-Bl{\'a}zquez}, P., {Vazdekis}, A.,
  {et~al.} 2011, \aap, 532, A95

\bibitem[{{Feldmeier-Krause} {et~al.}(2015){Feldmeier-Krause}, {Neumayer},
  {Sch{\"o}del}, {Seth}, {Hilker}, {de Zeeuw}, {Kuntschner}, {Walcher},
  {L{\"u}tzgendorf}, \& {Kissler-Patig}}]{Feldmeier-Krause2015}
{Feldmeier-Krause}, A., {Neumayer}, N., {Sch{\"o}del}, R., {et~al.} 2015, \aap,
  584, A2

\bibitem[{{Fellhauer} \& {Kroupa}(2002)}]{Fellhauer2002}
{Fellhauer}, M. \& {Kroupa}, P. 2002, \mnras, 330, 642

\bibitem[{{Fellhauer} \& {Kroupa}(2005)}]{Fellhauer2005}
{Fellhauer}, M. \& {Kroupa}, P. 2005, \mnras, 359, 223

\bibitem[{{Ferrarese} {et~al.}(2012){Ferrarese}, {C{\^o}t{\'e}}, {Cuilland re},
  {Gwyn}, {Peng}, {MacArthur}, {Duc}, {Boselli}, {Mei}, {Erben}, {McConnachie},
  {Durrell}, {Mihos}, {Jord{\'a}n}, {Lan{\c{c}}on}, {Puzia}, {Emsellem},
  {Balogh}, {Blakeslee}, {van Waerbeke}, {Gavazzi}, {Vollmer}, {Kavelaars},
  {Woods}, {Ball}, {Boissier}, {Courteau}, {Ferriere}, {Gavazzi},
  {Hildebrandt}, {Hudelot}, {Huertas-Company}, {Liu}, {McLaughlin}, {Mellier},
  {Milkeraitis}, {Schade}, {Balkowski}, {Bournaud}, {Carlberg}, {Chapman},
  {Hoekstra}, {Peng}, {Sawicki}, {Simard}, {Taylor}, {Tully}, {van Driel},
  {Wilson}, {Burdullis}, {Mahoney}, \& {Manset}}]{Ferrarese2012}
{Ferrarese}, L., {C{\^o}t{\'e}}, P., {Cuilland re}, J.-C., {et~al.} 2012,
  \apjs, 200, 4

\bibitem[{Forbes {et~al.}(2018)Forbes, Read, Gieles, \& Collins}]{Forbes2018}
Forbes, D.~A., Read, J.~I., Gieles, M., \& Collins, M. L.~M. 2018, MNRAS, 481,
  5592

\bibitem[{{Freudling} {et~al.}(2013){Freudling}, {Romaniello}, {Bramich},
  {Ballester}, {Forchi}, {Garc{\'{\i}}a-Dabl{\'o}}, {Moehler}, \&
  {Neeser}}]{Freudling2013}
{Freudling}, W., {Romaniello}, M., {Bramich}, D.~M., {et~al.} 2013, \aap, 559,
  A96

\bibitem[{{Georgiev} \& {B{\"o}ker}(2014)}]{Georgiev2014}
{Georgiev}, I.~Y. \& {B{\"o}ker}, T. 2014, \mnras, 441, 3570

\bibitem[{{Georgiev} {et~al.}(2016){Georgiev}, {B{\"o}ker}, {Leigh},
  {L{\"u}tzgendorf}, \& {Neumayer}}]{Georgiev2016}
{Georgiev}, I.~Y., {B{\"o}ker}, T., {Leigh}, N., {L{\"u}tzgendorf}, N., \&
  {Neumayer}, N. 2016, \mnras, 457, 2122

\bibitem[{{Georgiev} {et~al.}(2012){Georgiev}, {Goudfrooij}, \&
  {Puzia}}]{Georgiev2012}
{Georgiev}, I.~Y., {Goudfrooij}, P., \& {Puzia}, T.~H. 2012, \mnras, 420, 1317

\bibitem[{{Georgiev} {et~al.}(2008){Georgiev}, {Goudfrooij}, {Puzia}, \&
  {Hilker}}]{Georgiev2008}
{Georgiev}, I.~Y., {Goudfrooij}, P., {Puzia}, T.~H., \& {Hilker}, M. 2008, \aj,
  135, 1858

\bibitem[{{Georgiev} {et~al.}(2009{\natexlab{a}}){Georgiev}, {Hilker}, {Puzia},
  {Goudfrooij}, \& {Baumgardt}}]{Georgiev2009b}
{Georgiev}, I.~Y., {Hilker}, M., {Puzia}, T.~H., {Goudfrooij}, P., \&
  {Baumgardt}, H. 2009{\natexlab{a}}, \mnras, 396, 1075

\bibitem[{{Georgiev} {et~al.}(2010){Georgiev}, {Puzia}, {Goudfrooij}, \&
  {Hilker}}]{Georgiev2010}
{Georgiev}, I.~Y., {Puzia}, T.~H., {Goudfrooij}, P., \& {Hilker}, M. 2010,
  \mnras, 406, 1967

\bibitem[{{Georgiev} {et~al.}(2009{\natexlab{b}}){Georgiev}, {Puzia}, {Hilker},
  \& {Goudfrooij}}]{Georgiev2009a}
{Georgiev}, I.~Y., {Puzia}, T.~H., {Hilker}, M., \& {Goudfrooij}, P.
  2009{\natexlab{b}}, \mnras, 392, 879

\bibitem[{{Goerdt} {et~al.}(2006){Goerdt}, {Moore}, {Read}, {Stadel}, \&
  {Zemp}}]{Goerdt2006}
{Goerdt}, T., {Moore}, B., {Read}, J.~I., {Stadel}, J., \& {Zemp}, M. 2006,
  \mnras, 368, 1073

\bibitem[{{Guillard} {et~al.}(2016){Guillard}, {Emsellem}, \&
  {Renaud}}]{Guillard2016}
{Guillard}, N., {Emsellem}, E., \& {Renaud}, F. 2016, \mnras, 461, 3620

\bibitem[{{Ha{\c s}egan} {et~al.}(2005){Ha{\c s}egan}, {Jord{\'a}n},
  {C{\^o}t{\'e}}, {Djorgovski}, {McLaughlin}, {Blakeslee}, {Mei}, {West},
  {Peng}, {Ferrarese}, {Milosavljevi{\'c}}, {Tonry}, \&
  {Merritt}}]{Hasegan2005}
{Ha{\c s}egan}, M., {Jord{\'a}n}, A., {C{\^o}t{\'e}}, P., {et~al.} 2005, \apj,
  627, 203

\bibitem[{{Hanuschik} {et~al.}(2017){Hanuschik}, {Data Processing}, \& {Quality
  Control Group}}]{Hanuschik2017}
{Hanuschik}, R., {Data Processing}, \& {Quality Control Group}. 2017, in ESO
  Calibration Workshop: The Second Generation VLT Instruments and Friends, 15

\bibitem[{{Harris} \& {van den Bergh}(1981)}]{Harris1981}
{Harris}, W.~E. \& {van den Bergh}, S. 1981, \aj, 86, 1627

\bibitem[{{Hartmann} {et~al.}(2011){Hartmann}, {Debattista}, {Seth},
  {Cappellari}, \& {Quinn}}]{Hartmann2011}
{Hartmann}, M., {Debattista}, V.~P., {Seth}, A., {Cappellari}, M., \& {Quinn},
  T.~R. 2011, \mnras, 418, 2697

\bibitem[{{Herrmann} {et~al.}(2008){Herrmann}, {Ciardullo}, {Feldmeier}, \&
  {Vinciguerra}}]{Herrmann2008}
{Herrmann}, K.~A., {Ciardullo}, R., {Feldmeier}, J.~J., \& {Vinciguerra}, M.
  2008, \apj, 683, 630

\bibitem[{{Hilker} {et~al.}(2007){Hilker}, {Baumgardt}, {Infante},
  {Drinkwater}, {Evstigneeva}, \& {Gregg}}]{Hilker2007}
{Hilker}, M., {Baumgardt}, H., {Infante}, L., {et~al.} 2007, \aap, 463, 119

\bibitem[{{Hilker} \& {Richtler}(2000)}]{Hilker2000}
{Hilker}, M. \& {Richtler}, T. 2000, \aap, 362, 895

\bibitem[{{Ibata} {et~al.}(2019){Ibata}, {Bellazzini}, {Malhan}, {Martin}, \&
  {Bianchini}}]{Ibata2019}
{Ibata}, R.~A., {Bellazzini}, M., {Malhan}, K., {Martin}, N., \& {Bianchini},
  P. 2019, Nature Astronomy, 3, 667

\bibitem[{{Ibata} {et~al.}(1997){Ibata}, {Wyse}, {Gilmore}, {Irwin}, \&
  {Suntzeff}}]{Ibata1997}
{Ibata}, R.~A., {Wyse}, R. F.~G., {Gilmore}, G., {Irwin}, M.~J., \& {Suntzeff},
  N.~B. 1997, \aj, 113, 634

\bibitem[{{Ivkovich} \& {McCall}(2019)}]{Ivkovich2019}
{Ivkovich}, N. \& {McCall}, M.~L. 2019, \mnras, 486, 1964

\bibitem[{{Jerjen} {et~al.}(2000{\natexlab{a}}){Jerjen}, {Binggeli}, \&
  {Freeman}}]{Jerjen2000}
{Jerjen}, H., {Binggeli}, B., \& {Freeman}, K.~C. 2000{\natexlab{a}}, \aj, 119,
  593

\bibitem[{{Jerjen} {et~al.}(2000{\natexlab{b}}){Jerjen}, {Freeman}, \&
  {Binggeli}}]{Jerjen2000b}
{Jerjen}, H., {Freeman}, K.~C., \& {Binggeli}, B. 2000{\natexlab{b}}, \aj, 119,
  166

\bibitem[{{Johnson} {et~al.}(2015){Johnson}, {Rich}, {Pilachowski}, {Caldwell},
  {Mateo}, {Bailey}, \& {Crane}}]{Johnson2015}
{Johnson}, C.~I., {Rich}, R.~M., {Pilachowski}, C.~A., {et~al.} 2015, \aj, 150,
  63

\bibitem[{{Kacharov} {et~al.}(2018){Kacharov}, {Neumayer}, {Seth},
  {Cappellari}, {McDermid}, {Walcher}, \& {B{\"o}ker}}]{Kacharov2018}
{Kacharov}, N., {Neumayer}, N., {Seth}, A.~C., {et~al.} 2018, \mnras, 480, 1973

\bibitem[{{Karachentsev} {et~al.}(2007){Karachentsev}, {Tully}, {Dolphin},
  {Sharina}, {Makarova}, {Makarov}, {Sakai}, {Shaya}, {Kashibadze},
  {Karachentseva}, \& {Rizzi}}]{Karachentsev2007}
{Karachentsev}, I.~D., {Tully}, R.~B., {Dolphin}, A., {et~al.} 2007, \aj, 133,
  504

\bibitem[{{King}(1962)}]{King1962}
{King}, I. 1962, \aj, 67, 471

\bibitem[{{King} {et~al.}(2012){King}, {Bedin}, {Cassisi}, {Milone}, {Bellini},
  {Piotto}, {Anderson}, {Pietrinferni}, \& {Cordier}}]{King2012}
{King}, I.~R., {Bedin}, L.~R., {Cassisi}, S., {et~al.} 2012, \aj, 144, 5

\bibitem[{{Kirby} {et~al.}(2013){Kirby}, {Cohen}, {Guhathakurta}, {Cheng},
  {Bullock}, \& {Gallazzi}}]{Kirby2013}
{Kirby}, E.~N., {Cohen}, J.~G., {Guhathakurta}, P., {et~al.} 2013, \apj, 779,
  102

\bibitem[{{Kissler-Patig} {et~al.}(2006){Kissler-Patig}, {Jord{\'a}n}, \&
  {Bastian}}]{Kissler-Patig2006}
{Kissler-Patig}, M., {Jord{\'a}n}, A., \& {Bastian}, N. 2006, \aap, 448, 1031

\bibitem[{{Kovalev} {et~al.}(2018){Kovalev}, {Brinkmann}, {Bergemann}, \& {MPIA
  IT-department}}]{NLTE_MPIA}
{Kovalev}, M., {Brinkmann}, S., {Bergemann}, M., \& {MPIA IT-department}. 2018,
  {NLTE MPIA web server, [Online]. Available: {{http://nlte.mpia.de}} Max
  Planck Institute for Astronomy, Heidelberg.}

\bibitem[{{Lamers} {et~al.}(2017){Lamers}, {Kruijssen}, {Bastian}, {Rejkuba},
  {Hilker}, \& {Kissler-Patig}}]{Lamers2017}
{Lamers}, H.~J.~G.~L.~M., {Kruijssen}, J.~M.~D., {Bastian}, N., {et~al.} 2017,
  \aap, 606, A85

\bibitem[{{Lee} {et~al.}(2007){Lee}, {Gim}, \& {Casetti-Dinescu}}]{Lee2007}
{Lee}, Y.-W., {Gim}, H.~B., \& {Casetti-Dinescu}, D.~I. 2007, \apjl, 661, L49

\bibitem[{{Leung} {et~al.}(2019){Leung}, {Leaman}, {van de Ven}, \&
  {Battaglia}}]{Leung2019}
{Leung}, G. Y.~C., {Leaman}, R., {van de Ven}, G., \& {Battaglia}, G. 2019,
  arXiv e-prints, arXiv:1911.09167

\bibitem[{{Liu} {et~al.}(2015){Liu}, {Peng}, {C{\^o}t{\'e}}, {Ferrarese},
  {Jord{\'a}n}, {Mihos}, {Zhang}, {Mu{\~n}oz}, {Puzia}, {Lan{\c{c}}on}, {Gwyn},
  {Cuilland re}, {Blakeslee}, {Boselli}, {Durrell}, {Duc}, {Guhathakurta},
  {MacArthur}, {Mei}, {S{\'a}nchez-Janssen}, \& {Xu}}]{Liu2015}
{Liu}, C., {Peng}, E.~W., {C{\^o}t{\'e}}, P., {et~al.} 2015, \apj, 812, 34

\bibitem[{{Lotz} {et~al.}(2004){Lotz}, {Miller}, \& {Ferguson}}]{Lotz2004}
{Lotz}, J.~M., {Miller}, B.~W., \& {Ferguson}, H.~C. 2004, \apj, 613, 262

\bibitem[{{Lyubenova} \& {Tsatsi}(2019)}]{Lyubenova2019}
{Lyubenova}, M. \& {Tsatsi}, A. 2019, arXiv e-prints, arXiv:1903.10918

\bibitem[{{Lyubenova} {et~al.}(2013){Lyubenova}, {van den Bosch},
  {C{\^o}t{\'e}}, {Kuntschner}, {van de Ven}, {Ferrarese}, {Jord{\'a}n},
  {Infante}, \& {Peng}}]{Lyubenova2013}
{Lyubenova}, M., {van den Bosch}, R.~C.~E., {C{\^o}t{\'e}}, P., {et~al.} 2013,
  \mnras, 431, 3364

\bibitem[{{Maraston} {et~al.}(2004){Maraston}, {Bastian}, {Saglia},
  {Kissler-Patig}, {Schweizer}, \& {Goudfrooij}}]{Maraston2004}
{Maraston}, C., {Bastian}, N., {Saglia}, R.~P., {et~al.} 2004, \aap, 416, 467

\bibitem[{{Marino} {et~al.}(2015){Marino}, {Milone}, {Karakas}, {Casagrand e},
  {Yong}, {Shingles}, {Da Costa}, {Norris}, {Stetson}, {Lind}, {Asplund},
  {Collet}, {Jerjen}, {Sbordone}, {Aparicio}, \& {Cassisi}}]{Marino2015}
{Marino}, A.~F., {Milone}, A.~P., {Karakas}, A.~I., {et~al.} 2015, \mnras, 450,
  815

\bibitem[{{Mateo}(1998)}]{Mateo1998}
{Mateo}, M.~L. 1998, \araa, 36, 435

\bibitem[{{McGaugh} {et~al.}(2017){McGaugh}, {Schombert}, \&
  {Lelli}}]{McGaugh2017}
{McGaugh}, S.~S., {Schombert}, J.~M., \& {Lelli}, F. 2017, \apj, 851, 22

\bibitem[{{Meadows} {et~al.}(2019){Meadows}, {Navarro}, {Santos-Santos},
  {Benitez-Llambay}, \& {Frenk}}]{Meadows2019}
{Meadows}, N., {Navarro}, J.~F., {Santos-Santos}, I., {Benitez-Llambay}, A., \&
  {Frenk}, C. 2019, arXiv e-prints, arXiv:1910.11887

\bibitem[{{Mieske} {et~al.}(2013){Mieske}, {Frank}, {Baumgardt},
  {L{\"u}tzgendorf}, {Neumayer}, \& {Hilker}}]{Mieske2013}
{Mieske}, S., {Frank}, M.~J., {Baumgardt}, H., {et~al.} 2013, \aap, 558, A14

\bibitem[{{Mieske} {et~al.}(2002){Mieske}, {Hilker}, \& {Infante}}]{Mieske2002}
{Mieske}, S., {Hilker}, M., \& {Infante}, L. 2002, \aap, 383, 823

\bibitem[{{Mieske} {et~al.}(2004){Mieske}, {Hilker}, \& {Infante}}]{Mieske2004}
{Mieske}, S., {Hilker}, M., \& {Infante}, L. 2004, \aap, 418, 445

\bibitem[{{Mieske} {et~al.}(2008){Mieske}, {Hilker}, {Jord{\'a}n}, {Infante},
  {Kissler-Patig}, {Rejkuba}, {Richtler}, {C{\^o}t{\'e}}, {Baumgardt}, {West},
  {Ferrarese}, \& {Peng}}]{Mieske2008}
{Mieske}, S., {Hilker}, M., {Jord{\'a}n}, A., {et~al.} 2008, \aap, 487, 921

\bibitem[{{Mieske} \& {Kroupa}(2008)}]{MieskeKroupa2008}
{Mieske}, S. \& {Kroupa}, P. 2008, \apj, 677, 276

\bibitem[{{Mihos} \& {Hernquist}(1994)}]{MihosHernquist1994}
{Mihos}, J.~C. \& {Hernquist}, L. 1994, \apjl, 437, L47

\bibitem[{{Milone} {et~al.}(2019){Milone}, {Marino}, {Da Costa}, {Lagioia},
  {D'Antona}, {Goudfrooij}, {Jerjen}, {Massari}, {Renzini}, {Yong},
  {Baumgardt}, {Cordoni}, {Dondoglio}, {Li}, {Tailo}, {Asa'd}, \&
  {Ventura}}]{Milone2019}
{Milone}, A.~P., {Marino}, A.~F., {Da Costa}, G.~S., {et~al.} 2019, \mnras,
  2602

\bibitem[{{Milosavljevi{\'c}}(2004)}]{Milosavljevic2004}
{Milosavljevi{\'c}}, M. 2004, \apj, 605, L13

\bibitem[{{Misgeld} \& {Hilker}(2011)}]{Misgeld2011}
{Misgeld}, I. \& {Hilker}, M. 2011, \mnras, 414, 3699

\bibitem[{{M{\"u}ller} {et~al.}(2015){M{\"u}ller}, {Jerjen}, \&
  {Binggeli}}]{Muller2015}
{M{\"u}ller}, O., {Jerjen}, H., \& {Binggeli}, B. 2015, \aap, 583, A79

\bibitem[{{M{\"u}ller} {et~al.}(2017){M{\"u}ller}, {Jerjen}, \&
  {Binggeli}}]{Muller2017}
{M{\"u}ller}, O., {Jerjen}, H., \& {Binggeli}, B. 2017, \aap, 597, A7

\bibitem[{{M{\"u}ller} {et~al.}(2018{\natexlab{a}}){M{\"u}ller}, {Jerjen}, \&
  {Binggeli}}]{Mueller2018a}
{M{\"u}ller}, O., {Jerjen}, H., \& {Binggeli}, B. 2018{\natexlab{a}}, \aap,
  615, A105

\bibitem[{{M{\"u}ller} {et~al.}(2016){M{\"u}ller}, {Jerjen}, {Pawlowski}, \&
  {Binggeli}}]{Mueller2016}
{M{\"u}ller}, O., {Jerjen}, H., {Pawlowski}, M.~S., \& {Binggeli}, B. 2016,
  \aap, 595, A119

\bibitem[{{M{\"u}ller} {et~al.}(2018{\natexlab{b}}){M{\"u}ller}, {Pawlowski},
  {Jerjen}, \& {Lelli}}]{Mueller2018Sci}
{M{\"u}ller}, O., {Pawlowski}, M.~S., {Jerjen}, H., \& {Lelli}, F.
  2018{\natexlab{b}}, Science, 359, 534

\bibitem[{{M{\"u}ller} {et~al.}(2018{\natexlab{c}}){M{\"u}ller}, {Rejkuba}, \&
  {Jerjen}}]{Mueller2018}
{M{\"u}ller}, O., {Rejkuba}, M., \& {Jerjen}, H. 2018{\natexlab{c}}, \aap, 615,
  A96

\bibitem[{{M{\"u}ller} {et~al.}(2019){M{\"u}ller}, {Rejkuba}, {Pawlowski},
  {Ibata}, {Lelli}, {Hilker}, \& {Jerjen}}]{Mueller2019}
{M{\"u}ller}, O., {Rejkuba}, M., {Pawlowski}, M.~S., {et~al.} 2019, \aap, 629,
  A18

\bibitem[{{Murray}(2009)}]{Murray2009}
{Murray}, N. 2009, \apj, 691, 946

\bibitem[{{Norris} {et~al.}(2015){Norris}, {Escudero}, {Faifer}, {Kannappan},
  {Forte}, \& {van den Bosch}}]{Norris2015}
{Norris}, M.~A., {Escudero}, C.~G., {Faifer}, F.~R., {et~al.} 2015, \mnras,
  451, 3615

\bibitem[{{Ocvirk}(2010)}]{Ocvirk2010}
{Ocvirk}, P. 2010, \apj, 709, 88

\bibitem[{{Ordenes-Brice{\~n}o} {et~al.}(2018){Ordenes-Brice{\~n}o}, {Puzia},
  {Eigenthaler}, {Taylor}, {Mu{\~n}oz}, {Zhang}, {Alamo-Mart{\'{\i}}nez},
  {Ribbeck}, {Grebel}, {{\'A}ngel}, {C{\^o}t{\'e}}, {Ferrarese}, {Hilker},
  {Lan{\c c}on}, {Mieske}, {Miller}, {Rong}, \&
  {S{\'a}nchez-Janssen}}]{Ordenes2018}
{Ordenes-Brice{\~n}o}, Y., {Puzia}, T.~H., {Eigenthaler}, P., {et~al.} 2018,
  \apj, 860, 4

\bibitem[{{Paudel} {et~al.}(2011){Paudel}, {Lisker}, \&
  {Kuntschner}}]{Paudel2011}
{Paudel}, S., {Lisker}, T., \& {Kuntschner}, H. 2011, \mnras, 413, 1764

\bibitem[{{Perets} \& {Mastrobuono-Battisti}(2014)}]{Perets2014}
{Perets}, H.~B. \& {Mastrobuono-Battisti}, A. 2014, \apjl, 784, L44

\bibitem[{{Perina} {et~al.}(2011){Perina}, {Galleti}, {Fusi Pecci},
  {Bellazzini}, {Federici}, \& {Buzzoni}}]{Perina2011}
{Perina}, S., {Galleti}, S., {Fusi Pecci}, F., {et~al.} 2011, \aap, 531, A155

\bibitem[{{Pfeffer} \& {Baumgardt}(2013)}]{Pfeffer2013}
{Pfeffer}, J. \& {Baumgardt}, H. 2013, \mnras, 433, 1997

\bibitem[{{Pfeffer} {et~al.}(2014){Pfeffer}, {Griffen}, {Baumgardt}, \&
  {Hilker}}]{Pfeffer2014}
{Pfeffer}, J., {Griffen}, B.~F., {Baumgardt}, H., \& {Hilker}, M. 2014, \mnras,
  444, 3670

\bibitem[{{Phillipps} {et~al.}(2001){Phillipps}, {Drinkwater}, {Gregg}, \&
  {Jones}}]{Phillipps2001}
{Phillipps}, S., {Drinkwater}, M.~J., {Gregg}, M.~D., \& {Jones}, J.~B. 2001,
  \apj, 560, 201

\bibitem[{{Pietrinferni} {et~al.}(2004){Pietrinferni}, {Cassisi}, {Salaris}, \&
  {Castelli}}]{Pietrinferni2004}
{Pietrinferni}, A., {Cassisi}, S., {Salaris}, M., \& {Castelli}, F. 2004, \apj,
  612, 168

\bibitem[{{Pietrinferni} {et~al.}(2006){Pietrinferni}, {Cassisi}, {Salaris}, \&
  {Castelli}}]{Pietrinferni2006}
{Pietrinferni}, A., {Cassisi}, S., {Salaris}, M., \& {Castelli}, F. 2006, \apj,
  642, 797

\bibitem[{Pinna {et~al.}(2019)Pinna, Falc{\'{o}}n-Barroso, Martig, Sarzi,
  Coccato, Iodice, Corsini, de~Zeeuw, Gadotti, Leaman, Lyubenova, McDermid,
  Minchev, Morelli, van~de Ven, \& Viaene}]{Pinna2019}
Pinna, F., Falc{\'{o}}n-Barroso, J., Martig, M., {et~al.} 2019, Astronomy {\&}
  Astrophysics, 623, A19

\bibitem[{{Price-Whelan} {et~al.}(2018){Price-Whelan}, {Sip{\H{o}}cz},
  {G{\"u}nther}, {Lim}, {Crawford}, {Conseil}, {Shupe}, {Craig}, {Dencheva},
  {Ginsburg}, {VanderPlas}, {Bradley}, {P{\'e}rez-Su{\'a}rez}, {de Val-Borro},
  {Paper Contributors}, {Aldcroft}, {Cruz}, {Robitaille}, {Tollerud},
  {Coordination Committee}, {Ardelean}, {Babej}, {Bach}, {Bachetti}, {Bakanov},
  {Bamford}, {Barentsen}, {Barmby}, {Baumbach}, {Berry}, {Biscani}, {Boquien},
  {Bostroem}, {Bouma}, {Brammer}, {Bray}, {Breytenbach}, {Buddelmeijer},
  {Burke}, {Calderone}, {Cano Rodr{\'\i}guez}, {Cara}, {Cardoso}, {Cheedella},
  {Copin}, {Corrales}, {Crichton}, {D{\textquoteright}Avella}, {Deil},
  {Depagne}, {Dietrich}, {Donath}, {Droettboom}, {Earl}, {Erben}, {Fabbro},
  {Ferreira}, {Finethy}, {Fox}, {Garrison}, {Gibbons}, {Goldstein}, {Gommers},
  {Greco}, {Greenfield}, {Groener}, {Grollier}, {Hagen}, {Hirst}, {Homeier},
  {Horton}, {Hosseinzadeh}, {Hu}, {Hunkeler}, {Ivezi{\'c}}, {Jain}, {Jenness},
  {Kanarek}, {Kendrew}, {Kern}, {Kerzendorf}, {Khvalko}, {King}, {Kirkby},
  {Kulkarni}, {Kumar}, {Lee}, {Lenz}, {Littlefair}, {Ma}, {Macleod},
  {Mastropietro}, {McCully}, {Montagnac}, {Morris}, {Mueller}, {Mumford},
  {Muna}, {Murphy}, {Nelson}, {Nguyen}, {Ninan}, {N{\"o}the}, {Ogaz}, {Oh},
  {Parejko}, {Parley}, {Pascual}, {Patil}, {Patil}, {Plunkett}, {Prochaska},
  {Rastogi}, {Reddy Janga}, {Sabater}, {Sakurikar}, {Seifert}, {Sherbert},
  {Sherwood-Taylor}, {Shih}, {Sick}, {Silbiger}, {Singanamalla}, {Singer},
  {Sladen}, {Sooley}, {Sornarajah}, {Streicher}, {Teuben}, {Thomas},
  {Tremblay}, {Turner}, {Terr{\'o}n}, {van Kerkwijk}, {de la Vega}, {Watkins},
  {Weaver}, {Whitmore}, {Woillez}, {Zabalza}, \& {Contributors}}]{astropy:2018}
{Price-Whelan}, A.~M., {Sip{\H{o}}cz}, B.~M., {G{\"u}nther}, H.~M., {et~al.}
  2018, \aj, 156, 123

\bibitem[{{Puzia} {et~al.}(2014){Puzia}, {Paolillo}, {Goudfrooij}, {Maccarone},
  {Fabbiano}, \& {Angelini}}]{Puzia2014}
{Puzia}, T.~H., {Paolillo}, M., {Goudfrooij}, P., {et~al.} 2014, \apj, 786, 78

\bibitem[{{Puzia} \& {Sharina}(2008)}]{PuziaSharina2008}
{Puzia}, T.~H. \& {Sharina}, M.~E. 2008, \apj, 674, 909

\bibitem[{{Radburn-Smith} {et~al.}(2011){Radburn-Smith}, {de Jong}, {Seth},
  {Bailin}, {Bell}, {Brown}, {Bullock}, {Courteau}, {Dalcanton}, {Ferguson},
  {Goudfrooij}, {Holfeltz}, {Holwerda}, {Purcell}, {Sick}, {Streich}, {Vlajic},
  \& {Zucker}}]{Radburn-Smith2011}
{Radburn-Smith}, D.~J., {de Jong}, R.~S., {Seth}, A.~C., {et~al.} 2011, \apjs,
  195, 18

\bibitem[{{Recio-Blanco} {et~al.}(2006){Recio-Blanco}, {Aparicio}, {Piotto},
  {de Angeli}, \& {Djorgovski}}]{RecioBlanco2006}
{Recio-Blanco}, A., {Aparicio}, A., {Piotto}, G., {de Angeli}, F., \&
  {Djorgovski}, S.~G. 2006, \aap, 452, 875

\bibitem[{{Rejkuba}(2004)}]{Rejkuba2004}
{Rejkuba}, M. 2004, \aap, 413, 903

\bibitem[{{Rejkuba} {et~al.}(2007){Rejkuba}, {Dubath}, {Minniti}, \&
  {Meylan}}]{Rejkuba2007}
{Rejkuba}, M., {Dubath}, P., {Minniti}, D., \& {Meylan}, G. 2007, \aap, 469,
  147

\bibitem[{{Rossa} {et~al.}(2006){Rossa}, {van der Marel}, {B{\"o}ker},
  {Gerssen}, {Ho}, {Rix}, {Shields}, \& {Walcher}}]{Rossa2006}
{Rossa}, J., {van der Marel}, R.~P., {B{\"o}ker}, T., {et~al.} 2006, \aj, 132,
  1074

\bibitem[{{S{\'a}nchez-Janssen}
  {et~al.}(2019{\natexlab{a}}){S{\'a}nchez-Janssen}, {C{\^o}t{\'e}},
  {Ferrarese}, {Peng}, {Roediger}, {Blakeslee}, {Emsellem}, {Puzia},
  {Spengler}, {Taylor}, {{\'A}lamo-Mart{\'\i}nez}, {Boselli}, {Cantiello},
  {Cuillandre}, {Duc}, {Durrell}, {Gwyn}, {MacArthur}, {Lan{\c{c}}on}, {Lim},
  {Liu}, {Mei}, {Miller}, {Mu{\~n}oz}, {Mihos}, {Paudel}, {Powalka}, \&
  {Toloba}}]{SanchezJanssen2019}
{S{\'a}nchez-Janssen}, R., {C{\^o}t{\'e}}, P., {Ferrarese}, L., {et~al.}
  2019{\natexlab{a}}, \apj, 878, 18

\bibitem[{{S{\'a}nchez-Janssen}
  {et~al.}(2019{\natexlab{b}}){S{\'a}nchez-Janssen}, {Puzia}, {Ferrarese},
  {C{\^o}t{\'e}}, {Eigenthaler}, {Miller}, {Ordenes-Brice{\~n}o}, {Peng},
  {Ribbeck}, {Roediger}, {Spengler}, \& {Taylor}}]{SanchezJanssen2019b}
{S{\'a}nchez-Janssen}, R., {Puzia}, T.~H., {Ferrarese}, L., {et~al.}
  2019{\natexlab{b}}, \mnras, 486, L1

\bibitem[{{Schiavon} {et~al.}(2004){Schiavon}, {Rose}, {Courteau}, \&
  {MacArthur}}]{Schiavon2004}
{Schiavon}, R.~P., {Rose}, J.~A., {Courteau}, S., \& {MacArthur}, L.~A. 2004,
  \apjl, 608, L33

\bibitem[{{Schinnerer} {et~al.}(2008){Schinnerer}, {B{\"o}ker}, {Meier}, \&
  {Calzetti}}]{Schinnerer2008}
{Schinnerer}, E., {B{\"o}ker}, T., {Meier}, D.~S., \& {Calzetti}, D. 2008,
  \apjl, 684, L21

\bibitem[{{Schlafly} \& {Finkbeiner}(2011)}]{Schlafly2011}
{Schlafly}, E.~F. \& {Finkbeiner}, D.~P. 2011, \apj, 737, 103

\bibitem[{{Seth} {et~al.}(2010){Seth}, {Cappellari}, {Neumayer}, {Caldwell},
  {Bastian}, {Olsen}, {Blum}, {Debattista}, {McDermid}, {Puzia}, \&
  {Stephens}}]{Seth2010}
{Seth}, A.~C., {Cappellari}, M., {Neumayer}, N., {et~al.} 2010, \apj, 714, 713

\bibitem[{{Seth} {et~al.}(2014){Seth}, {van den Bosch}, {Mieske}, {Baumgardt},
  {Brok}, {Strader}, {Neumayer}, {Chilingarian}, {Hilker}, {McDermid},
  {Spitler}, {Brodie}, {Frank}, \& {Walsh}}]{Seth2014}
{Seth}, A.~C., {van den Bosch}, R., {Mieske}, S., {et~al.} 2014, \nat, 513, 398

\bibitem[{{Sharina} {et~al.}(2008){Sharina}, {Karachentsev}, {Dolphin},
  {Karachentseva}, {Tully}, {Karataeva}, {Makarov}, {Makarova}, {Sakai},
  {Shaya}, {Nikolaev}, \& {Kuznetsov}}]{Sharina2008}
{Sharina}, M.~E., {Karachentsev}, I.~D., {Dolphin}, A.~E., {et~al.} 2008,
  \mnras, 384, 1544

\bibitem[{{Sills} {et~al.}(2019){Sills}, {Dalessandro}, {Cadelano},
  {Alfaro-Cuello}, \& {Kruijssen}}]{Sills2019}
{Sills}, A., {Dalessandro}, E., {Cadelano}, M., {Alfaro-Cuello}, M., \&
  {Kruijssen}, J.~M.~D. 2019, \mnras, 490, L67

\bibitem[{{Soto} {et~al.}(2016){Soto}, {Lilly}, {Bacon}, {Richard}, \&
  {Conseil}}]{Soto2016}
{Soto}, K.~T., {Lilly}, S.~J., {Bacon}, R., {Richard}, J., \& {Conseil}, S.
  2016, \mnras, 458, 3210

\bibitem[{{Spekkens} {et~al.}(2014){Spekkens}, {Urbancic}, {Mason}, {Willman},
  \& {Aguirre}}]{Spekkens2014}
{Spekkens}, K., {Urbancic}, N., {Mason}, B.~S., {Willman}, B., \& {Aguirre},
  J.~E. 2014, \apjl, 795, L5

\bibitem[{{Spengler} {et~al.}(2017){Spengler}, {C{\^o}t{\'e}}, {Roediger},
  {Ferrarese}, {S{\'a}nchez-Janssen}, {Toloba}, {Liu}, {Guhathakurta},
  {Cuillandre}, {Gwyn}, {Zirm}, {Mu{\~n}oz}, {Puzia}, {Lan{\c{c}}on}, {Peng},
  {Mei}, \& {Powalka}}]{Spengler2017}
{Spengler}, C., {C{\^o}t{\'e}}, P., {Roediger}, J., {et~al.} 2017, \apj, 849,
  55

\bibitem[{{Stetson}(1987)}]{Stetson1987}
{Stetson}, P.~B. 1987, \pasp, 99, 191

\bibitem[{{Strader} {et~al.}(2013){Strader}, {Seth}, {Forbes}, {Fabbiano},
  {Romanowsky}, {Brodie}, {Conroy}, {Caldwell}, {Pota}, {Usher}, \&
  {Arnold}}]{Strader2013}
{Strader}, J., {Seth}, A.~C., {Forbes}, D.~A., {et~al.} 2013, \apjl, 775, L6

\bibitem[{{Taylor} {et~al.}(2018){Taylor}, {Eigenthaler}, {Puzia}, {Mu{\~n}oz},
  {Ribbeck}, {Zhang}, {Ordenes-Brice{\~n}o}, \& {Bovill}}]{Taylor2018}
{Taylor}, M.~A., {Eigenthaler}, P., {Puzia}, T.~H., {et~al.} 2018, \apjl, 867,
  L15

\bibitem[{{Taylor} {et~al.}(2016){Taylor}, {Mu{\~n}oz}, {Puzia}, {Mieske},
  {Eigenthaler}, \& {Bovill}}]{Taylor2016}
{Taylor}, M.~A., {Mu{\~n}oz}, R.~P., {Puzia}, T.~H., {et~al.} 2016,
  arxiv:1608.07285

\bibitem[{{Taylor} {et~al.}(2010){Taylor}, {Puzia}, {Harris}, {Harris},
  {Kissler-Patig}, \& {Hilker}}]{Taylor2010}
{Taylor}, M.~A., {Puzia}, T.~H., {Harris}, G.~L., {et~al.} 2010, \apj, 712,
  1191

\bibitem[{{Tolstoy} {et~al.}(2009){Tolstoy}, {Hill}, \& {Tosi}}]{Tolstoy2009}
{Tolstoy}, E., {Hill}, V., \& {Tosi}, M. 2009, \araa, 47, 371

\bibitem[{{Tonry} \& {Davis}(1979)}]{tonry1979}
{Tonry}, J. \& {Davis}, M. 1979, \aj, 84, 1511

\bibitem[{{Tremaine} {et~al.}(1975){Tremaine}, {Ostriker}, \&
  {Spitzer}}]{Tremaine1975}
{Tremaine}, S.~D., {Ostriker}, J.~P., \& {Spitzer}, Jr., L. 1975, \apj, 196,
  407

\bibitem[{{Tully} {et~al.}(2015){Tully}, {Libeskind}, {Karachentsev},
  {Karachentseva}, {Rizzi}, \& {Shaya}}]{Tully2015}
{Tully}, R.~B., {Libeskind}, N.~I., {Karachentsev}, I.~D., {et~al.} 2015,
  \apjl, 802, L25

\bibitem[{{Turner} {et~al.}(2012){Turner}, {C{\^o}t{\'e}}, {Ferrarese},
  {Jord{\'a}n}, {Blakeslee}, {Mei}, {Peng}, \& {West}}]{Turner2012}
{Turner}, M.~L., {C{\^o}t{\'e}}, P., {Ferrarese}, L., {et~al.} 2012, \apjs,
  203, 5

\bibitem[{{Usher} {et~al.}(2019){Usher}, {Brodie}, {Forbes}, {Romanowsky},
  {Strader}, {Pfeffer}, \& {Bastian}}]{Usher2019}
{Usher}, C., {Brodie}, J.~P., {Forbes}, D.~A., {et~al.} 2019, arXiv e-prints,
  arXiv:1909.05753

\bibitem[{{Vazdekis} {et~al.}(1996){Vazdekis}, {Casuso}, {Peletier}, \&
  {Beckman}}]{Vazdekis1996}
{Vazdekis}, A., {Casuso}, E., {Peletier}, R.~F., \& {Beckman}, J.~E. 1996,
  \apjs, 106, 307

\bibitem[{{Vazdekis} {et~al.}(2010){Vazdekis}, {S{\'a}nchez-Bl{\'a}zquez},
  {Falc{\'o}n-Barroso}, {Cenarro}, {Beasley}, {Cardiel}, {Gorgas}, \&
  {Peletier}}]{Vazdekis2010}
{Vazdekis}, A., {S{\'a}nchez-Bl{\'a}zquez}, P., {Falc{\'o}n-Barroso}, J.,
  {et~al.} 2010, \mnras, 404, 1639

\bibitem[{{Voggel} {et~al.}(2016){Voggel}, {Hilker}, \&
  {Richtler}}]{Voggel2016}
{Voggel}, K., {Hilker}, M., \& {Richtler}, T. 2016, \aap, 586, A102

\bibitem[{{Walcher} {et~al.}(2006){Walcher}, {B{\"o}ker}, {Charlot}, {Ho},
  {Rix}, {Rossa}, {Shields}, \& {van der Marel}}]{Walcher2006}
{Walcher}, C.~J., {B{\"o}ker}, T., {Charlot}, S., {et~al.} 2006, \apj, 649, 692

\bibitem[{{Walcher} {et~al.}(2005){Walcher}, {van der Marel}, {McLaughlin},
  {Rix}, {B{\"o}ker}, {H{\"a}ring}, {Ho}, {Sarzi}, \& {Shields}}]{Walcher2005}
{Walcher}, C.~J., {van der Marel}, R.~P., {McLaughlin}, D., {et~al.} 2005,
  \apj, 618, 237

\bibitem[{{Warren} {et~al.}(2004){Warren}, {Jerjen}, \&
  {Koribalski}}]{Warren2004}
{Warren}, B.~E., {Jerjen}, H., \& {Koribalski}, B.~S. 2004, \aj, 128, 1152

\bibitem[{{Warren} {et~al.}(2006){Warren}, {Jerjen}, \&
  {Koribalski}}]{Warren2006}
{Warren}, B.~E., {Jerjen}, H., \& {Koribalski}, B.~S. 2006, \aj, 131, 2056

\bibitem[{{Weilbacher} {et~al.}(2012){Weilbacher}, {Streicher}, {Urrutia},
  {Jarno}, {P{\'e}contal-Rousset}, {Bacon}, \& {B{\"o}hm}}]{Weilbacher2012}
{Weilbacher}, P.~M., {Streicher}, O., {Urrutia}, T., {et~al.} 2012, in
  \procspie, Vol. 8451, Software and Cyberinfrastructure for Astronomy II,
  84510B

\bibitem[{{Woodley}(2006)}]{Woodley2006}
{Woodley}, K.~A. 2006, \aj, 132, 2424

\bibitem[{{Woodley} {et~al.}(2010){Woodley}, {G{\'o}mez}, {Harris}, {Geisler},
  \& {Harris}}]{Woodley2010}
{Woodley}, K.~A., {G{\'o}mez}, M., {Harris}, W.~E., {Geisler}, D., \& {Harris},
  G. L.~H. 2010, \aj, 139, 1871

\bibitem[{{Woodley} {et~al.}(2007){Woodley}, {Harris}, {Beasley}, {Peng},
  {Bridges}, {Forbes}, \& {Harris}}]{Woodley2007}
{Woodley}, K.~A., {Harris}, W.~E., {Beasley}, M.~A., {et~al.} 2007, \aj, 134,
  494

\bibitem[{{Zinnecker} {et~al.}(1988){Zinnecker}, {Keable}, {Dunlop}, {Cannon},
  \& {Griffiths}}]{Zinnecker1988}
{Zinnecker}, H., {Keable}, C.~J., {Dunlop}, J.~S., {Cannon}, R.~D., \&
  {Griffiths}, W.~K. 1988, in IAU Symposium, Vol. 126, The Harlow-Shapley
  Symposium on Globular Cluster Systems in Galaxies, ed. J.~E. {Grindlay} \&
  A.~G.~D. {Philip}, 603

\end{thebibliography}

\end{document}